\documentclass[12pt,preprint]{aastex}

\shorttitle{Short Title} \shortauthors{Fan}

\begin{document}

%% LaTeX will automatically break titles if they run longer than
%% one line. However, you may use \\ to force a line break if
%% you desire.

\title{Noisy weak-lensing convergence peak statistics near clusters of galaxies and beyond}

%% Use \author, \affil, and the \and command to format
%% author and affiliation information.
%% Note that \email has replaced the old \authoremail command
%% from AASTeX v4.0. You can use \email to mark an email address
%% anywhere in the paper, not just in the front matter.
%% As in the title, you can use \\ to force line breaks.

\author{Zuhui Fan, Huanyuan Shan and Jiayi Liu}
\affil{Department of Astronomy, Peking University,
    Beijing 100871, China}
\email{fan@bac.pku.edu.cn}

%% Notice that each of these authors has alternate affiliations, which
%% are identified by the \altaffilmark after each name.  Specify alternate
%% affiliation information with \altaffiltext, with one command per each
%% affiliation.

%\altaffiltext{1}{Visiting Astronomer, Cerro Tololo Inter-American Observatory.
%CTIO is operated by AURA, Inc.\ under contract to the National Science
%Foundation.}
%\altaffiltext{2}{Society of Fellows, Harvard University.}
%\altaffiltext{3}{present address: Center for Astrophysics,
%    60 Garden Street, Cambridge, MA 02138}
%\altaffiltext{4}{Visiting Programmer, Space Telescope Science Institute}
%\altaffiltext{5}{Patron, Alonso's Bar and Grill}

%% Mark off your abstract in the ``abstract'' environment. In the manuscript
%% style, abstract will output a Received/Accepted line after the
%% title and affiliation information. No date will appear since the author
%% does not have this information. The dates will be filled in by the
%% editorial office after submission.

%\begin{abstract}
%This is a preliminary report on surface photometry of the major
%fraction of known globular clusters, to see which of them show the signs
%of a collapsed core.
%We also explore some diversionary mathematics and recreational tables.
%\end{abstract}
\begin{abstract}

Taking into account noise from intrinsic ellipticities of source galaxies, 
in this paper, we study the peak statistics in weak-lensing convergence maps
around clusters of galaxies and beyond. We emphasize how the noise peak
statistics is affected by the density distribution of nearby clusters, 
and also how cluster-peak signals are changed by the existence of noise.
These are the important aspects to be understood thoroughly in weak-lensing analyses 
for individual clusters as well as in cosmological applications of weak-lensing cluster statistics. 
We adopt Gaussian smoothing with the smoothing scale $\theta_G=0.5\hbox{ arcmin}$ in our analyses.
It is found that the noise peak distribution near a cluster of galaxies
depends sensitively on the density profile of the cluster. 
For a cored isothermal cluster with the core radius $R_c$, 
the inner region with $R\le R_c$ appears noisy containing  
on average $\sim 2.4$ peaks with $\nu\ge 5$ for $R_c= 1.7\hbox{ arcmin}$
and the true peak height of the cluster $\nu=5.6$, where $\nu$ denotes 
the convergence signal to noise ratio. 
For a NFW cluster of the same mass and the same central $\nu$, 
the average number of peaks with $\nu\ge 5$ within $R\le R_c$ is $\sim 1.6$. 
Thus a high peak corresponding to the main cluster can be identified 
more cleanly in the NFW case.
%For a NFW cluster with the characteristic scale $R_s$ and the true central $\nu$ 
%the same as those of the isothermal case, a well-defined single high peak corresponding 
%to the main cluster can be commonly seen. 
In the outer region with $R_c<R\le 5R_c$,  
the number of high noise peaks is considerably enhanced in comparison with 
that of the pure noise case without the nearby cluster. 
For $\nu\ge 4$, depending on the treatment of the mass-sheet degeneracy in weak-lensing analyses,
the enhancement factor $f$ is in the range
of $\sim 5$ to $\sim 55$ for both clusters as their outer density profiles are
similar. 
%For $\nu\ge 4$, the enhancement factor $f$ is in the range
%of $\sim 5$ to $\sim 50$ and $\sim 1$ to $\sim 3$ for the isothermal and the NFW clusters,
%respectively, depending on the treatment of the mass-sheet degeneracy in weak-lensing analyses. 
The properties of the main-cluster-peak identified in convergence maps
are also affected significantly by the presence of noise. 
Scatters as well as a systematic shift for the peak height show up. 
The height distribution is peaked at $\nu\sim 6.6$, rather than at $\nu= 5.6$, 
corresponding to a shift of $\Delta \nu \sim 1$, for the isothermal cluster. 
For the NFW cluster, $\Delta \nu \sim 0.8$.
The existence of noise also causes a location offset for the weak-lensing identified 
main-cluster-peak with respect to the true center of the cluster.
The offset distribution is very broad and extends to $R\sim R_c$ for the isothermal case.
For the NFW cluster, it is relatively narrow and peaked at $R\sim 0.2R_c$. 
We also analyze NFW clusters of different concentrations.
It is found that the more centrally concentrated the mass distribution of a cluster is, the less
its weak-lensing signal is affected by noise.
Incorporating these important effects and the mass function of NFW dark matter halos, 
we further present a model calculating the statistical abundances of total convergence peaks, 
true and false ones, over a large field beyond individual clusters. 
The results are in good agreement with those from numerical simulations.  
The model then allows us to probe cosmologies with the convergence peaks directly without 
the need of expensive follow-up observations to differentiate true and false peaks. 

\end{abstract}

%% Keywords should appear after the \end{abstract} command. The uncommented
%% example has been keyed in ApJ style. See the instructions to authors
%% for the journal to which you are submitting your paper to determine
%% what keyword punctuation is appropriate.


%% From the front matter, we move on to the body of the paper.
%% In the first two sections, notice the use of the natbib \citep

%%k_theta_1_note.txt and \citet commands to identify citations.  The citations are
%% tied to the reference list via symbolic KEYs. The KEY corresponds
%% to the KEY in the \bibitem in the reference list below. We have
%% chosen the first three characters of the first author's name plus
%% the last two numeral of the year of publication as our KEY for
%% each reference.

%\section{Introduction}
%
%A focal problem today in the dynamics of globular clusters is
%core collapse.  It has been predicted by theory
%for decades \citep{hen61,lyn68,spi85}, but
%observation has been less alert to the phenomenon. For many years the
%central brightness peak in M15 \citep{kin75,new78}
%seemed a unique anomaly.  Then \citet{aur82} suggested a central peak
%in NGC 6397, and a limited photographic survey of ours \citep[Paper I]{djo84}
%found three more cases, including NGC 6624, whose
%sharp center had often been remarked on \citep{can78}.
%ii
%i$
%i$
\section{Introduction}

Arising from the light deflection by the inhomogeneous mass distribution of the universe, 
gravitational weak-lensing effects have been applied to map out, individually or statistically, 
the mass distribution of galaxies, clusters of galaxies, and large-scale
structures (e.g., Bartelmann \& Schneider 2001; Mandelbaum et al. 2008; Li et al. 2009; 
Clowe et al. 2006; Corless et al. 2009; Fu et al. 2008). Deep observations also 
reveal the time evolution of the structure formation (Massey et al. 2007).
Cosmic shear analyses have been used to constrain different cosmological parameters 
(e.g., Benjamin et al. 2007; Kilbinger 2009; Li et al. 2009). Future wide and deep weak-lensing surveys
will give rise to observational data with greatly improved quantity and quality.
Therefore the derived cosmological information is expected to be much more accurate 
than that of today, which will make the weak-lensing effect one of the most important probes 
in cosmology, notably, in understanding the nature of dark energy (e.g., Albrecht et al.
2006). However, it has been realized that weak-lensing effects are not as clean as they 
appear to be. Being very weak, they can be affected and 
contaminated easily by different observational and physical effects. 
Considerable efforts have been paid to understand these effects and their influences on 
cosmological studies (e.g., Tang \& Fan 2005; Heymans et al. 2006; Ma et al. 2006; 
Huterer et al. 2006; Bridle \& King 2007; Fan 2007; Amara \& Refregier 2008; Sun et al. 2009).  

Among others, intrinsic ellipticities of source galaxies are known 
to dominate over their weak-lensing induced shear signals by a large factor. The 
environmental dependence of galaxy formation can lead to intrinsic-intrinsic 
and shear-intrinsic correlations, which can contaminate directly the two-point weak-lensing 
shear correlations at percent to tens of percent level (e.g., Hirata \& Seljak 2004; 
Mandelbaum et al. 2006; Fan 2007; Zhang 2008; Joachimi \& Schneider 2009).
Moreover, in reconstructing the foreground mass distribution from shear measurements of
background galaxies, their intrinsic ellipticities leave considerable noise 
even if they are completely random (e.g., van Waerbeke 2000; Hamana et al. 2004; White et al. 2002). 
%Different statistical methods have been developed to suppress such noises in 
%the weak-lensing mass reconstruction, and smoothing is one of the simplest ways for doing so.
%Given the rms of the intrinsic ellipticities, the amplitude of the smoothed noise 
%depends on the smoothing scale and on the surface number density of source galaxies 
%thus the observational depth (e.g., van Waerbeke 2000). 
Understanding fully the statistical properties of the noise field
and the mutual influence between noise and the true lensing effects are therefore
crucially important in studies aiming to reveal the mass distribution of large-scale structures.
Particularly, chance alignments of intrinsic ellipticities can form false peaks in a 
lensing convergence map, and mimic 'dark clumps', namely structures with large mass to light ratios
(e.g., Erben et al. 2000; Linden et al. 2006; Gavazzi \& Soucail 2007; Schirmer et al. 2007;
Fan 2007). Furthermore, noise can affect true lensing signals, causing large scatters
and possibly systematic biases to them. 
These can lead to considerable uncertainties in analyzing the mass distribution 
of clusters of galaxies, in investigating the possible existence of 'dark clumps', 
and in deriving constraints on cosmological parameters from weak-lensing cluster abundances.

Different statistical methods have been developed to suppress such noise in
weak-lensing mass reconstructions, and the smoothing is one of the simplest ways for doing so.
Given the rms of intrinsic ellipticities, the amplitude of the smoothed noise
depends on the smoothing scale and on the surface number density of source galaxies
thus the observational depth (e.g., van Waerbeke 2000).
%The statistical properties of the noise depend on the method applied to 
%derive the convergence from the shear measurements. For the smoothing methodology, 
Because of the central limit theorem, it is expected that the statistics of the noise field 
after smoothing should be close to Gaussian if a large enough number of galaxies are included 
within the smoothing window. By employing a Gaussian smoothing function, 
van Waerbeke (2000) shows that the noise field of the convergence 
is indeed approximately Gaussian.  He further analyzes its properties, such as
the number and the shape distribution of the noise peaks. Fan (2007) extends the study
by including the intrinsic alignment of background galaxies in the analyses, 
and shows that the number of false peaks are enhanced depending on the 
strength of the intrinsic alignment. Schneider (1996) consider the effects 
of noise on the height of the peaks associated with true mass concentrations
in aperture-mass maps by adding a Gaussian scatter to the peak height. 
From numerical simulations, Hamana et al. (2004) analyze noise effects on true peaks, 
and present a phenomenological correction to the abundance of weak-lensing detected peaks. 
In this paper, we put forward a theoretical model to study the mutual
influence of noise and true peaks. Specifically, we consider the lensing effects
of a cluster of galaxies taking into account noise from intrinsic ellipticities of source 
galaxies. We investigate the effect of the density profile of the 
cluster on the distribution of noise peaks, 
and conversely, how noise can change the true peak signal of the cluster. Based on these
analyses and the mass function of dark matter halos, we develop a model to compute 
statistically weak-lensing convergence peak abundances over a large field, 
including both the peaks associated with clusters of galaxies and the false ones 
from noise. Therefore we can use convergence peaks directly   
as cosmological probes without the need to sort out true and false peaks. 
Moreover, the distribution of noise peaks depends on the density profile of true clusters, 
and thus contains additional cosmological information, which is naturally included 
in our model. 

The rest of the paper is organized as follows. In \S2, we present our theoretical 
framework. The results are shown in \S3 with \S3.1 for individual cluster cases 
and \S3.2 for statistical results on a large field. \S4 contains summaries and discussions.

\section{Peak statistics in noisy convergence maps around clusters of galaxies}

With the intrinsic ones taken into account, the observed ellipticities of source galaxies 
can be written in the weak lensing regime as (e.g., Bartelmann \& Schneider 2001)
\begin{equation}\label{w1}
\mathbf{\epsilon^{(O)}}\approx {\mathbf{\gamma}} + \mathbf{\epsilon^{(S)}},
\end{equation}
where ${\mathbf{\gamma}}$ is the complex lensing shear due to the foreground mass 
distributions, and $\mathbf{\epsilon^{(O)}}$ and $\mathbf{\epsilon^{(S)}}$ represent the 
observed and the intrinsic complex ellipticities of source galaxies, respectively. 
The corresponding smoothed quantities are 
\begin{equation}\label{w1}
\mathbf{E}^{(O)}(\vec \theta)=\mathbf\Gamma(\vec \theta)
+{1\over n_g}\sum_{i=1}^{N_g} W(\vec \theta -\vec \theta_i)\mathbf{\epsilon}^{(S)}(\vec \theta_i),
\end{equation}
where $\mathbf{E}^{(O)}$ and $\mathbf{\Gamma}$ are
the smoothed $\mathbf{\epsilon^{(O)}}$ and $\gamma$, respectively,
$W(\vec \theta)$ is the smoothing function, and $n_g$ and $N_g$ are,
respectively, the surface number density and the total number of source galaxies in the field.

The lensing convergence $\kappa$ can be obtained from the shear through their 
following relation in the Fourier space 
\begin{equation}\label{w1}
\tilde\kappa(\vec k)=c_{\alpha}(k)\tilde \gamma_{\alpha}(\vec k),
\end{equation}
where the summation over $\alpha=(1,2)$ is implied, and
$c_{\alpha}=[\cos (2\varphi),\sin (2\varphi)]$ with
$\vec k=k(\cos \varphi,\sin \varphi)$ (Kaiser \& Squires 1993).
Then the real-space convergence reconstructed from the observed ellipticities is 
\begin{equation}\label{w1}
\kappa_n(\vec \theta)=\int d\vec k \hbox{ } e^{-i\vec k \cdot \vec \theta}
c_{\alpha}(k)\tilde \epsilon^{(O)}_{\alpha}(\vec k)=\kappa(\vec \theta)+n(\vec \theta),
\end{equation} 
where $\kappa$ and $n$ are the lensing convergence and the noise 
from the intrinsic ellipticities, respectively. After smoothing, we have 
\begin{equation}\label{w1}
K_N(\vec \theta)=K(\vec \theta)+N(\vec \theta)
\end{equation}
with $K$ and $N$ the corresponding smoothed $\kappa$ and $n$. 
It is therefore clear that in order to extract the true $K$ from the noisy $K_N$,
we must understand the influence of $N$.

In this paper, we mainly consider foreground clusters of galaxies as lens objects.
%Both individually and then statistically based on the results from individual studies 
%and the mass function of dark matter halos.  
%In this paper, we particularly study the noise effects near a cluster of galaxies,
Considering an individual cluster, we take $K(\vec \theta)$ to be the smoothed 
convergence of the cluster, which is determined by its assumed known density profile. 
The noise $N$ is regarded as a random field. The total $K_N(\vec \theta)$ is therefore 
also a random field with its statistics determined by the noise field 
$N(\vec \theta)$ and modulated by the cluster $K(\vec \theta)$. Concerning 
peaks in the noisy convergence $K_N(\vec \theta)$ around an individual cluster, 
there should be a dominant one corresponding to the main cluster. There are also false peaks
arising from $N$. It is expected while the properties of the dominant peak in $K_N(\vec \theta)$
should be closely related to those of true $K(\vec \theta)$, they are inevitably affected
by the existence of $N$. On the other hand, the noise peaks in $K_N(\vec \theta)$ 
should also be modulated by the cluster $K(\vec \theta)$ in comparison with 
those in pure noise field $N$. It should be noted that the peak statistics 
depend not only on the value of $K_N$, but also on its profile through the first
and second derivatives of $K_N$. Thus $K$ and $N$ need to $^\prime$collaborate$^\prime$ to give rise
to peaks in $K_N$. Therefore the peak properties in $K_N$ differ from those in $N$ or in $K$
in ways that are more complex than just changing the peak height by a value of $K$ or $N$.
Our main focus in the paper is to understand
such mutual influences between $K$ and $N$ in terms of the peak properties in $K_N$,
and further their effects on relevant cosmological studies. 
The formulations presented in the following are for noisy $K_N(\vec \theta)$ around 
a cluster with a known density distribution. These set the theoretical framework for 
analyses presented in \S 3.2 on large-scale statistics beyond individual clusters
incorporating the mass function of dark matter halos.
 
It has been shown that the smoothed noise field $N$ is approximately Gaussian  
because of the central limit theorem (van Waerbeke 2000; Fan 2007). 
For an individual cluster, its convergence $K$ is regarded as a known quantity. 
Thus the statistics of the field $K_N$ is also Gaussian and follows that of $N$
with the modifications due to the presence of $K$ from the cluster.  
For the random Gaussian field $K_N$, its peak statistics
involves random variables $K_N$, $\partial_i K_N$ (denoted as $K^i_{N}$) and $\partial_{ij} K_N$ 
(denoted as $K^{ij}_{N}$) ($i,j=1,2$).
Their joint probability distribution function can be readily written out as follows
(e.g., Bardeen et al. 1986; Bond \& Efstathiou 1987)
\begin{eqnarray}\label{w11}
&& p(K_N,K^{11}_{N},K^{22}_{N},K^{12}_{N},K^1_{N},K^2_{N})\hbox{\ }dK_N\hbox{\ }dK^{11}_{N}\hbox{\ }dK^{22}_{N}\hbox{ }dK^{12}_{N}\hbox{ }dK^1_{N}\hbox{ }dK^2_{N}= \nonumber \\
&& {1\over [2\pi(1-\gamma_N^2)\sigma_0]^{1/2}}\exp\bigg \{-{\{(K_N-K)/\sigma_0+\gamma_N [(K^{11}_N-K^{11})
+(K^{22}_N-K^{22})]/\sigma_2\}^2\over 2(1-\gamma_N^2)}\bigg \}\times \nonumber \\
&& {1\over 2\pi\sigma_2^2}\exp\bigg \{-{[(K^{11}_N-K^{11})-(K^{22}_N-K^{22})]^2\over 2 \sigma_2^2}
-{(K^{11}_N-K^{11})^2\over \sigma_2^2}-{(K^{22}_N-K^{22})^2\over \sigma_2^2}\bigg \}\times \nonumber \\
&& {8\over (2\pi)^{1/2}\sigma_2}\exp\bigg \{-{4(K^{12}_N-K^{12})^2\over \sigma_2^2}\bigg \}
\times \nonumber \\
&&{1\over \pi\sigma_1^2}\exp\bigg [-{(K^1_N-K^1)^2\over \sigma_1^2}-{(K^2_N-K^2)^2\over \sigma_1^2}\bigg ] \hbox{\ }dK_N\hbox{\ }dK^{11}_{N}\hbox{\ }dK^{22}_{N}\hbox{ }dK^{12}_{N}\hbox{ }dK^1_{N}\hbox{ }dK^2_{N},
\end{eqnarray}
where all the quantities related to $K_N$ are attached a subscript N, while those
without the subscript refer to the ones from $K$. 
The $\sigma_0$, $\sigma_1$ and $\sigma_2$ are the moments of the noise field $N$, which are 
given by, with $N(k)$ being the Fourier transform of $N$ (e.g., van Waerbeke 2000),
\begin{equation}\label{w1}
\sigma_i^2=\int {d\vec k}\hbox{ }k^{2i}<|N(k)|^2>.
\end{equation}
The quantity $\gamma_N$ is defined as $\gamma_N=\sigma_1^2/(\sigma_0\sigma_2)$. By diagonalizing the
second derivative tensor $- K^{ij}_N$ to obtain the eigen values of $\lambda_{N1}$
and $\lambda_{N2}$ and by defining 
\begin{equation}\label{w1}
x_N={\lambda_{N1}+\lambda_{N2}\over \sigma_2}, \quad e_N={(\lambda_{N1}-\lambda_{N2})\over 2\sigma_2x_N},
\end{equation}
we have 
\begin{equation}\label{w1}
K^{11}_N=-{\sigma_2x_N\over 2}(1+2e_N\cos 2\theta_N), \quad K^{22}_N=-{\sigma_2x_N\over 2}(1-2e_N\cos 2\theta_N), \quad K^{12}=-\sigma_2 x_N e_N\sin 2\theta_N,
\end{equation}
where $\theta_N$ is the rotation angle in the range of $[0, \pi]$.
We further define $\nu_N=K_N/\sigma_0$, then Eq. (6) can be simplified
to
\begin{eqnarray}\label{w11}
&& p(\nu_N,x_N,e_N,\theta_N,K^1_{N},K^2_{N})\hbox{\ }d\nu_N\hbox{\ }dx_N\hbox{\ }de_N
\hbox{ }d\theta_N\hbox{ }dK^1_{N}\hbox{ }dK^2_{N}= \nonumber \\
&& \exp\bigg [-{1\over 2}{(K^{11}-K^{22})^2\over \sigma_2^2}-{(K^{11})^2+(K^{22})^2\over \sigma_2^2}
-4{(K^{12})^2\over \sigma_2^2}\bigg ]\times \nonumber \\
&& \bigg \{{d\nu_N\over [2\pi(1-\gamma_N^2)]^{1/2}}\exp\bigg \{-{[(\nu_N-K/\sigma_0)-\gamma_N x_N
-\gamma_N(K^{11}+K^{22})/\sigma_2]^2\over 2(1-\gamma_N^2)}\bigg \}\times \nonumber \\
&& {dx_N\over (2\pi)^{1/2}}\exp\bigg (-{x_N^2\over 2}\bigg ){8\over \pi}x_N^2e_N\hbox{ }de_N\hbox{ }d\theta_N
\exp(-4x_N^2e_N^2)\exp\bigg [-x_N\bigg ({K^{11}+K^{22}\over \sigma_2}\bigg )\bigg ]\times \nonumber \\
&& \exp\bigg [-4x_Ne_N\cos2\theta_N\bigg ({K^{11}-K^{22}\over \sigma_2}\bigg )\bigg ]
\exp\bigg (-8x_Ne_N\sin2\theta_N{K^{12}\over \sigma_2}\bigg ) \times \nonumber \\
&&{1\over \pi\sigma_1^2}\exp\bigg [-{(K^1_N-K^1)^2\over \sigma_1^2}-{(K^2_N-K^2)^2\over \sigma_1^2}
\bigg ] \hbox{ }dK^1_{N}\hbox{ }dK^2_{N}\bigg \}.
\end{eqnarray}

Considering peak statistics, the average number density of maxima given $\nu_N=\nu_0, x_N=x_0, 
e_N=e_0$, and $\theta_N=\theta_0$ is (Bond \& Efstathiou 1987)
%\begin{equation}\label{w1}
\begin{eqnarray}\label{w11}
n_{peak}(\nu_0, x_0, e_0, \theta_0)=&&  
<\delta(\nu_N-\nu_0)\delta(x_N-x_0)\delta(e_N-e_0)\delta(\theta_N-\theta_0) \nonumber \\
&& \delta(K^1_N)\delta(K^1_N)(\sigma_2^2/4) x_N^2(1-4e_N^2)\Theta(1-2e_N)\Theta(e_N)>, 
\end{eqnarray}
%\end{equation}
where the average is calculated with the probability function given by Eq. (10), and
the step functions $\Theta(1-2e_N)$ and $\Theta(e_N)$ are required to satisfy the
conditions of maxima with $\lambda_{N1}\ge \lambda_{N2}\ge 0$.
%It is seen then that the number of peaks near a cluster of galaxies depends on the
%mass distribution of the cluster through $K$, $K^i$, and $K^{ij}$ $(i, j=1, 2)$.
Thus the number density of peaks given $\nu_N=\nu_0$ and $x_N=x_0$ is 
\begin{eqnarray}\label{w11}
&& n_{peak}(\nu_0, x_0)=
\exp\bigg [-{1\over 2}{(K^{11}-K^{22})^2\over \sigma_2^2}-{(K^{11})^2+(K^{22})^2\over \sigma_2^2}
-4{(K^{12})^2\over \sigma_2^2}\bigg ]\exp \bigg [-{(K^1)^2+(K^2)^2\over \sigma_1^2}\bigg ]
\times \nonumber \\
&& \bigg \{ {1\over 2\pi\theta_*^2}{1\over [2\pi(1-\gamma_N^2)]^{1/2}}\bigg\} 
\exp\bigg \{-{[(\nu_0-K/\sigma_0)-\gamma_N x_0
-\gamma_N(K^{11}+K^{22})/\sigma_2]^2\over 2(1-\gamma_N^2)}\bigg \}\times \nonumber \\
&& {1\over (2\pi)^{1/2}}\exp\bigg [-{x_0^2\over 2}-x_0\bigg ({K^{11}+K^{22}\over \sigma_2}\bigg )\bigg ]
\int_0^{1/2}de_N \hbox{ }{8\over \pi}(x_0^2e_N)x_0^2(1-4e_N^2) \exp(-4x_0^2e_N^2) \times \nonumber \\
&& \int_0^{\pi} d\theta_N \hbox{ }
\exp\bigg [-4x_0e_N\cos2\theta_N\bigg ({K^{11}-K^{22}\over \sigma_2}\bigg )\bigg ]
\exp\bigg (-8x_0e_N\sin2\theta_N{K^{12}\over \sigma_2}\bigg ), \nonumber \\
\end{eqnarray}
where $\theta_*^2=2\sigma_1^2/\sigma_2^2$. We reorganize the above equation and obtain the following
\begin{eqnarray}\label{w11}
n_{peak}(\nu_0, x_0)=&&\exp \bigg [-{(K^1)^2+(K^2)^2\over \sigma_1^2}\bigg ]
\bigg \{ {1\over 2\pi\theta_*^2}{1\over (2\pi)^{1/2}}\bigg\}
\exp\bigg [-{1\over 2}\bigg ( \nu_0-{K\over \sigma_0}\bigg )^2\bigg ] \times \nonumber \\
&&{1\over [2\pi(1-\gamma_N^2)]^{1/2}}\exp\bigg \{-{ [{x_0+(K^{11}+K^{22})/ \sigma_2
-\gamma_N(\nu_0-K/\sigma_0)}]^2\over 2(1-\gamma_N^2)}\bigg \} \times  F(x_0), \nonumber \\ 
\end{eqnarray}
where 
\begin{eqnarray}\label{w11}
F(x_0)=&& \exp\bigg [-\bigg ({K^{11}-K^{22}\over \sigma_2}\bigg )^2-4{(K^{12})^2\over \sigma_2^2}\bigg ]
\times \nonumber \\
&& \int_0^{1/2}de_N \hbox{ }8(x_0^2e_N)x^2(1-4e_N^2) \exp(-4x_0^2e_N^2) \times \nonumber \\
&& \int_0^{\pi} {d\theta_N\over \pi} \hbox{ }
\exp\bigg [-4x_0e_N\cos2\theta_N\bigg ({K^{11}-K^{22}\over \sigma_2}\bigg )\bigg ]\times
\nonumber \\
&& \exp\bigg (-8x_0e_N\sin2\theta_N{K^{12}\over \sigma_2}\bigg ).
\end{eqnarray}
Further we have
\begin{eqnarray}\label{w11}
n_{peak}(\nu_0)=&&\exp \bigg [-{(K^1)^2+(K^2)^2\over \sigma_1^2}\bigg ]
\bigg \{ {1\over 2\pi\theta_*^2}{1\over (2\pi)^{1/2}}\bigg\}
\exp\bigg [-{1\over 2}\bigg ( \nu_0-{K\over \sigma_0}\bigg )^2\bigg ] \times \nonumber \\
&&\int {dx_N\over [2\pi(1-\gamma_N^2)]^{1/2}}\exp\bigg \{-{ [{x_N+(K^{11}+K^{22})/ \sigma_2
-\gamma_N(\nu_0-K/\sigma_0)}]^2\over 2(1-\gamma_N^2)}\bigg \} \times  F(x_N). \nonumber \\
\end{eqnarray}

It is seen clearly that the number of peaks near a cluster of galaxies depends on the
density profile of the cluster through $K$, $K^i$, and $K^{ij}$ $(i, j=1, 2)$.
Assuming a spherical mass distribution for a cluster of galaxies, we can calculate
$n_{peak}(\nu_0)$ at different radii along a specific direction 
where $K^{12}=0$ in our chosen coordinate system. Then we have
\begin{eqnarray}\label{w11}
F(x_N)=&& \exp\bigg [-{(K^{11}-K^{22})^2\over \sigma_2^2}\bigg ]
\times \nonumber \\
&& \int_0^{1/2}de_N \hbox{ }8(x_N^2e_N)x_N^2(1-4e_N^2) \exp(-4x_N^2e_N^2)
\times \nonumber \\
&& \int_0^{\pi} {d\theta_N\over \pi} \hbox{ }
\exp\bigg [-4x_Ne_N\cos 2\theta_N{(K^{11}-K^{22})\over \sigma_2}\bigg ]. \nonumber \\
\end{eqnarray}

Without the foreground cluster, the function $F(x_N)$ can be written out analytically as
$F(x_N)=x_N^2-1+\exp(-x_N^2)$ (Bond \& Efstathiou 1987). 
With a non-zero $K^{11}-K^{22}$, we find a fitting formula for $F(x_N)$ with
\begin{equation}\label{w1}
F(x_N)=x_N^2-1+\exp(-x_N^2)+C_1^2\{\exp[-x_N^3/g(C_1)]-1\},
\end{equation}
where $C_1=(K^{11}-K^{22})/\sigma_2$ and $g(C_1)$ is a fitted value depending on $C_1$.
In Figure 1, we show two sets of results for $F(x_N)-[x_N^2-1+\exp(-x_N^2)]$ 
with $C_1=0.2$ (upper set) and $C_1=0.5$ (lower set), respectively.
For each set, both the numerical result (solid line) and the one calculated from the fitting formula 
(dashed line) are shown. For $C_1=0.2$ and $C_1=0.5$, we have $g(C_1)=3.3$ and $g(C_1)=3.5$, 
respectively. It is seen that Eq.(17) provides an excellent approximation to $F(x_N)$,
which will be adopted for further calculations.

Given a spherical cluster profile, we can obtain its convergence $\kappa$, and further the
smoothed $K$ with a smoothing function $W$ through
\begin{equation}\label{w1}
K(\vec x)=\int \hbox{ }d\vec {x^{'}} \hbox{ }W(\vec {x^{'}}-\vec x)\kappa(\vec {x^{'}}).
\end{equation}
The relevant derivatives are 
\begin{equation}\label{w1}
K^i=\partial_iK(\vec x)=\int \hbox{ }d\vec {x^{'}} \hbox{ }\partial_iW(\vec {x^{'}}-\vec x)\kappa(\vec {x^{'}}),
\end{equation}
and
\begin{equation}\label{w1}
K^{ij}=\partial_{ij}K(\vec x)=\int \hbox{ }d\vec {x^{'}} \hbox{ }\partial_{ij}W(\vec {x^{'}}-\vec x)\kappa(\vec {x^{'}}).
\end{equation}

In our analyese, we use the Gaussian smoothing function with 
\begin{equation}\label{w1}
W(\vec x)={1\over \pi \theta_G^2}\exp\bigg (-{|\vec x|^2\over \theta_G^2}\bigg ),
\end{equation}
where $\theta_G$ is the smoothing scale. We calculate $K$, $K^i$ and 
$K^{ij}$ along the $x_1$ (i.e, $x_2=0$) direction. Then the relevant
quantities become
\begin{equation}\label{w1}
K(x_1, 0)=\int \hbox{ }dx_1^{'}dx_2^{'} \hbox{ }{1\over \pi \theta_G^2}
\exp\bigg [-{(x_1^{'}-x_1)^2+(x_2^{'})^2\over \theta_G^2}\bigg ]\kappa(x_1^{'}, x_2^{'}),
\end{equation}

\begin{equation}\label{w1}
K^1(x_1, x_2)|_{x_2=0}=\int \hbox{ }dx_1^{'}dx_2^{'} \hbox{ }{1\over \pi \theta_G^2}
{2\over \theta_G^2}(x_1^{'}-x_1)
\exp\bigg [-{(x_1^{'}-x_1)^2+(x_2^{'})^2\over \theta_G^2}\bigg ]\kappa(x_1^{'}, x_2^{'}),
\end{equation}

\begin{eqnarray}\label{w11}
&&[K^{11}(x_1, x_2)+K^{22}(x_1, x_2)]|_{x_2=0}= 
\int \hbox{ }dx_1^{'}dx_2^{'} \hbox{ }{1\over \pi \theta_G^2}
{(-4)\over \theta_G^2} \exp\bigg [-{(x_1^{'}-x_1)^2+(x_2^{'})^2\over \theta_G^2}\bigg ]
\kappa(x_1^{'}, x_2^{'}) \nonumber \\
&& + \int \hbox{ }dx_1^{'}dx_2^{'} \hbox{ }{1\over \pi \theta_G^2}
{4\over \theta_G^2}{(x_1^{'}-x_1)^2+(x_2^{'})^2\over \theta_G^2} \exp\bigg [-{(x_1^{'}-x_1)^2+(x_2^{'})^2\over \theta_G^2}\bigg ]\kappa(x_1^{'},x_2^{'}), 
\end{eqnarray}

\begin{eqnarray}\label{w11}
[K^{11}(x_1, x_2)-K^{22}(x_1, x_2)]|_{x_2=0}=\int dx_1^{'}dx_2^{'} &&
{1\over \pi \theta_G^2} {4\over \theta_G^2}{(x_1^{'}-x_1)^2-(x_2^{'})^2\over \theta_G^2}\nonumber \\
&& \times \exp\bigg [-{(x_1^{'}-x_1)^2+(x_2^{'})^2\over \theta_G^2}\bigg ]\kappa(x_1^{'},x_2^{'}),
\nonumber \\
\end{eqnarray}
and $K^2=0$ and $K^{12}=0$.

We consider two types of most commonly used cluster profiles, namely,
the isothermal profile and the NFW profile (Navarro, Frenk \& White 1996).
For the isothermal one, we adopt the non-singular isothermal sphere model with its lensing potential 
having the form (van Waerbeke 2000)
\begin{equation}\label{w1}
\phi(x_1,x_2)=A_{\phi}\sqrt{1+{(x_1-x_{10})^2+(x_2-x_{20})^2\over R_c^2}},
\end{equation}
where $R_c$ is the core radius, $(x_{10},x_{20})$ is the central position of the cluster,
and 
\begin{equation}\label{w1}
A_{\phi}=(4\pi\sigma_{c}^2/c^2)(D_{ol}D_{ls}/D_{os})R_c^2.
\end{equation}
Here $D_{ol}, D_{ls}$ and $D_{os}$ are the angular diameter distances 
between the observer and the lens, between the 
lens and the source, and between the observer and the source, respectively, and 
$\sigma_{c}$ represents the velocity dispersion of the isothermal cluster.
We choose $(x_{10},x_{20})=(0,0)$, then the convergence $\kappa$ of the cluster is
\begin{equation}\label{w1}
\kappa(x_1,x_2)={\kappa_0\over 2}{2+(x_1^2+x_2^2)/R_c^2\over [1+(x_1^2+x_2^2)/R_c^2]^{3/2}},
\end{equation}
where $\kappa_0$ is related to $A_{\phi}$ and $R_c$. 

For the NFW profile, we have (Navarro, Frenk \& White 1996)
\begin{equation}\label{w1}
\rho(r)={\rho_s\over r/r_s(1+r/r_s)^2},
\end{equation}
where $\rho_s$ and $r_s$ are the characteristic mass density and scale of a dark matter
halo. The corresponding convergence is $\kappa(R)=(2\rho_sr_s/\Sigma_c)f(R/r_s)$ with
\begin{eqnarray}\label{w11}
f(\tilde R^s)&& =-{1\over \bigg [1-{(\tilde R^s)}^2\bigg ]}+{1\over \bigg [1-{(\tilde R^s)}^2\bigg]^{3/2}}\hbox{ arccosh}
(1/ {\tilde R^s})\quad \quad \hbox{for } \tilde R^s <1
\nonumber \\ &&
={1\over 3} \qquad \qquad \qquad \qquad \qquad \qquad \quad \quad \qquad \qquad \qquad 
\quad \hbox{for } \tilde R^s=1 \nonumber \\ &&
=-{1\over \bigg [1-{(\tilde R^s)}^2\bigg ]}-{1\over \bigg [{(\tilde R^s)}^2-1\bigg ]^{3/2}}\hbox{ arccos}(1/ {\tilde R^s})\quad \quad \hbox{for } \tilde R^s >1
\end{eqnarray}
where $\tilde R^s=\sqrt{x_1^2+x_2^2}/r_s$. Given the redshift of the lens, 
$\tilde R^s$ can also be expressed in angular coordinates with 
$\tilde R^s=\sqrt{(x_1/D_l)^2+(x_2/D_l)^2}/(r_s/D_l)$, so for the isothermal case of Eq. (28).

In Figure 2, we show the smoothed quantities $K/\sigma_0$, $K^1/\sigma_1$,
$(K^{11}+K^{22})/\sigma_2$, and $(K^{11}-K^{22})/\sigma_2$ for
the two profiles, respectively. The specific parameters are
$\theta_G=0.5\hbox{ arcmin}$, $R_c=1.7\hbox{ arcmin}$ and $\kappa_0/\sigma_0=6$
for the isothermal model (solid lines). For the NFW model (dashed lines), 
we adjust $R_s$ and the central amplitude 
of $\kappa(R)$ so that it has the same mass and the same smoothed central $K$ 
as those of the isothermal cluster. In this case, $R_s=5R_c$.
Note that the horizontal axes in the plots are $\tilde R=R/R_c$.
It is seen that the two profiles are different, especially in the inner region.
Thus we expect to see differences of the peak statistics in the two cases.  

%It is also noted that with a Gaussian smoothing, we have
%\begin{equation}\label{w1}
%\sigma^2_{0}={\sigma^2_{\epsilon}\over 2}{1\over 2\pi\theta_G^2 n_g},
%\quad \sigma_1=\sqrt{2}\sigma_0/\theta_G, \quad \sigma_2=2\sqrt{2}\sigma_0/\theta_G^2.
%\end{equation}
%It is noted that except $K/\sigma_0$, all the other three quantities are much less than one,
%at $\tilde R>2$ for both models. Therefore, we expect that in the region with $\tilde R>2$, 
%the effect of the mass distribution of the cluster on noise peak statistics comes in mainly through
%$K/\sigma_0$. This is in analogous to the peak-background split scenario in explaining the biased 
%halo distributions with respect to the mass density distribution. In the weak lensing case, 
%however, the mass-sheet degeneracy can cause some complications, which we will touch upon 
%later in the paper.

\section{Results}

This section contains two parts. In \S 3.1, we show the results around individual clusters.
%and emphasize the influence of the density profile of a cluster on the peak statistics  
%in its weak lensing convergence map. 
The isothermal and the NFW clusters are analyzed. The field extends 
to $5R_c$ for all the considered cases. The analyses shown 
%to $5R_c$ and $5R_s$ for the two clusters, respectively. The analyses shown 
here are relevant to weak lensing studies targeted 
at individual clusters. In \S 3.2, the statistical results over a large field are shown.
Specifically, we present a theoretical model based on the mass function and the NFW density profile
of dark matter halos to predict the numbers of high peaks in the weak lensing convergence map 
over a large field. Our model takes into account the effects of noise on the peak heights of 
true clusters and the enhancement of noise peaks near clusters of galaxies.
Given a cosmology, our model can then predict the number distribution of peaks with different
heights. These cosmology-dependent quantities can be used as very promising and 
efficient probes to constrain cosmologies. The advantages of these probes are 
two folds. Firstly, because we concern the total number of peaks, we do not need to 
differentiate true peaks and false peaks, which often requires extensive follow-up observations. 
Secondly, the redshift information of the peaks is not necessary since we 
use the peak height distribution rather than the redshift distribution of peaks as
cosmological probes. 
%In this paper, we mainly present the model, and its cosmological applications
%will be investigated in details in our forthcoming paper. 

\subsection{Individual cluster fields}

Here we discuss the peak statistics around individual clusters of galaxies. Two aspects
are addressed. One is how the distribution of noise peaks is influenced by the density profile
of a cluster.  The other is how the properties of the peak corresponding to the true cluster,
such as its height and position, are affected by the presence of noise.
In our calculations, we assume that a considered cluster is located at the center, and the 
field extends to $5R_c$. Our main results are presented for the two clusters shown in Figure 2. 
In addition, we also analyze different NFW clusters of the same mass but with different
concentrations to further demonstrate the dependence of the peak statistics on cluster
profiles. Besides theoretical analyses, we perform Monte Carlo simulations
with different realizations of the noise field added to the considered cluster $\kappa$-map. The
theoretical results are then compared with the results averaged over $100$ simulations
for each of the clusters.

For a visual impression, we first present in Figure 3 a typical set of maps 
from our Monte Carlo simulations for the two clusters shown in Figure 2. Figure 3a and Figure 3b
are for the isothermal cluster, and the NFW cluster, respectively. It is seen that in the regions
outside $R_c$, there is almost a one-to-one correspondence between the peaks in 
the two maps. This is because in these outer regions, the density profiles of the two clusters
are relatively flat with small values of $K^1/\sigma_1$, $K^{11}+K^{22}/\sigma_2$ and
$K^{11}-K^{22}/\sigma_2$. Then the peaks are essentially the ones existed in the pure noise
field but with their heights modified by the term $K/\sigma_0$. However, it should be noted
that $K/\sigma_0$ depends on the radius $R$, and thus the peak statistics over a range of 
$R$ cannot be described by an overall constant shift in peak height from that of $N$. Within $R_c$, 
the peaks look very different in the two maps, which clearly shows the effects of the 
density profile ($K^1/\sigma_1$, $K^{11}+K^{22}/\sigma_2$ and $K^{11}-K^{22}/\sigma_2$) 
in addition to the term $K/\sigma_0$. For the NFW cluster, there is a single high peak
which corresponds clearly to the peak signal of the main cluster. For the isothermal cluster, however, 
two prominent peaks with comparable heights occur. The positions of the peaks are also
shifted from the center. Such peak configurations due to noise can cause 
complications in weak lensing studies of the structure of clusters. 

To be quantitative, we present in Figure 4
the radial distributions of the number of peaks above a certain 
threshold around the two clusters. In each panel, the lines are the results
from our theoretical calculations, and the symbols are the average results from the $100$ 
Monte Carlo simulations for each case. The dash-dotted line and the '+' symbols 
are for the NFW cluster. The solid line and the diamonds are for the isothermal cluster.  
The vertical axes are for the number of peaks within rings of width $\Delta \tilde R=0.1$,
where $\tilde R=R/R_c$.
Different scales of the horizontal and vertical axes in different panels should be noted. 
It is seen that the results from the theoretical model and from the simulations agree very well. 
The peak distributions for the two cluster profiles are distinctively different,
especially in the inner region and for high peaks with $\nu\ge 5$. 
For the NFW cluster, high peaks are mainly found near the 
central region of the cluster, and the 
total number of high peaks is smaller than that of isothermal cluster.
%exist mainly in the central region, and very few peaks 
%with $\nu>4$ can be found in regions with $\tilde R >1$. For the isothermal cluster, 
%however, high peaks can be seen over a broad region. There is a significant fraction 
%of peaks with $\nu>5$ appearing at $\tilde R >1$, causing complications in the 
%interpretations of $^\prime$dark clumps$^\prime$ near clusters of galaxies. 
In the region with $R < R_c$, the average numbers of peaks for 
$\nu\ge 3,\hbox{} 4, \hbox{} 5$ and $\hbox{} 6$
%are approximately $1.8$, $1.1$, $0.73$ and $0.37$, respectively, for the NFW case.
are approximately $3$, $2.6$, $1.7$ and $0.8$, respectively, for the NFW case.
For the isothermal case, the corresponding numbers are $3.5$, $3.2$, $2.4$ and $1.1$. Therefore,
for the NFW cluster, the inner region is relatively clean and often contains a single and well defined
high peak as seen in Figure 3b. On the other hand, for the isothermal cluster, 
there are on average $2$ peaks with $\nu>5$ in the inner region, 
of which, one corresponds to the real cluster and the other is generated due to noise (Figure 3a). 
 
We further consider the inner ($R\le R_c$) and outer ($R_c < R\le 5R_c$) regions separately. 
In the outer regions, we analyze the occurrence probability of
high noise peaks, which is relevant to the dark clump problem. 
We emphasize, with respect to the case without clusters,
the enhancement due to the cluster density distribution. 
% 
%In Figure 3, we show the average cumulative number of peaks in the cluster field for the isothermal
%(solid line) and NFW models (dash-dotted line), respectively. For comparison, the results 
%for a blank field of the same size without clusters are also shown (dashed line).
%We see that the number of peaks above a certain threshold is enhanced around a cluster, and
%the enhancement depends on the mass profile of the cluster. 
%
%The results shown in Figure 3 include the true peak associated with the known cluster.
%Concerning the contamination of noise peaks to the possible existence of true
%dark clumps around the known cluster, it is of more interest to consider noise peaks outside
%the central cluster region. 
Figure 5a shows the distribution of the cumulative number of peaks above the threshold $\nu_t$
in the region $1<\tilde R\le 5$ ($\tilde R=R/R_c$)
and Figure 5b presents the corresponding enhancement factor 
$f=N_{peak}(\nu_t)/N_{ran}(\nu_t)$, where $N_{peak}(\nu_t)$ and $N_{ran}(\nu_t)$ denote,
respectively, the cumulative numbers of peaks 
in the cases with and without a cluster in the center. In Figure 5a, the solid, dash-dotted 
and dashed lines are for the isothermal cluster, the NFW cluster, and the case without a 
cluster, respectively. The diamond symbols and the plus symbols are for the results averaged
over $100$ simulations of the isothermal and NFW clusters, respectively. In Figure 5b,
the solid and dash-dotted lines are for the corresponding enhancement factors of the
two clusters.
%To see how well our theoretical model works, we conduct $100$
%Monte Carlo simulations. 
%Each simulation has a field of view $10 R_c\times 10 R_c$, and is done
%by reconstructing $\kappa$ with the Kaiser-Squires method from the 'observed' shear of source galaxies.
%Each component of the 'observed' shear of a galaxy is a superposition of that from 
%the cluster calculated from the lensing potential given in Eq. (20) and that from 
%the intrinsic ellipticities modeled as Gaussian random variables. 
%All the relevant parameters are set to be the same as those in our theoretical analyses.
%The peak counting is then performed on each $\kappa$ map after a Gaussian smoothing. 
%The average number distribution of peaks from the $100$ simulated maps are shown 
%as triangles in Fig.4a. It is seen that the results from our theoretical model
%agree very well with those from simulations, which validates our theoretical treatments.
%For the enhancement factor $f$, we find that besides the dependence on the mass profile, it
%increases sensitively with the threshold $\nu_t$. 
%It is seen that the enhancement is much more significant for the 
%isothermal cluster case than that of the NFW cluster. 
It is seen that the results are similar for the two clusters because
their density profiles are similar in the outer regions as shown in Figure 2. 
However, differences can still be seen for high peaks. For $\nu_t>5$, 
the enhancement factor is larger for the isothermal cluster
than that for the NFW cluster. This is due to the fact that in the outer regions,
these high peaks are mainly found in $1<\tilde R<2$, where $K/\sigma_0$ is higher
for the isothermal cluster (Figure 2a).
For $\nu_{t}=3$, $4$, $5$, $6$, 
$f\sim 9.8$, $55$, $595$ and $10000$ for the isothermal cluster.
%The corresponding $f$ for the NFW cluster are $2$, $2.6$, $3.6$, $5.3$ and $5000$.
The corresponding $f$ for the NFW cluster are $10.7$, $55$, $437$, and $4671$.
Therefore depending on the density profile of a cluster, 
the occurrence probability of high noise peaks around it can be greatly boosted. The effects
are larger for higher peaks. 

%From Figure 2, it is seen that except $K/\sigma_0$, all the other three quantities related 
%to the mass distribution of the cluster are much less than one at $\tilde R>1$ for both models. 
As discussed earlier, in the outer region, the effect of the density profile of a cluster 
on noise peak statistics comes in mainly through the term $K/\sigma_0$. 
It increases the heights of the peaks originally with 
$\nu_{t}-K/\sigma_0<\nu<\nu_{t}$ to be above the threshold $\nu_{t}$. Because of the 
approximately exponential dependence of $N_{peak}(\nu)$ on $\nu$, the enhancement at 
$\nu_{t}\ge 3$ changes sensitively with the values of $K/\sigma_0$ and $\nu_{t}$. 
%Given the same central peak height and the characteristic scale,
%the NFW profile drops much faster, leading to lower $K/\sigma_0$
%and thus fewer noise peaks above the threshold than that of the isothermal cluster. 
This enhancement in outer regions can be analogous to the peak-background 
split scenario in explaining the biased
dark matter halo distributions with respect to the dark matter mass density distribution
(Bardeen et al. 1986; Mo \& White 1996). It is noted, however, that $K/\sigma_0$
changes with radius. Thus the local enhancement depends on the radius, and the 
overall enhancement factor $f$ over the outer region with $1<\tilde R\le 5$ cannot be
modeled by a constant shift of $\nu_{t}$ with respect to that of $N$. 
When the observational field is comparable to the size of a cluster, we should also be 
aware that the mass-sheet degeneracy in lensing analyses
brings some arbitrariness in the reconstructed mass distribution, 
which in turn affects the noise peak statistics within the field. 
In the weak lensing regime, this arbitrariness is mainly reflected by an additive constant to 
$\kappa$. In real studies, different people may choose different constants. It can be
chosen so that the averaged $<\kappa>=0$ in the observed field (e.g., Erben et al. 2000), or 
to make the $\kappa$ averaged over the outer-most ring of the observed field be zero 
(e.g., Kaiser \& Squires 2003).
It is also attempted to recover the true mass distribution of a cluster by adjusting
the constant to make the reconstructed mass distribution be consistent with the distribution 
determined by some other observations (e.g., Umetsu \& Futamase 2000). 
%It is clear from the above explanations that
%the number of noise peaks above a certain threshold is affected by the chosen constant. 
In Figure 6a and Figure 6b, we show $f$ for the isothermal and NFW clusters,
respectively, by applying an additive constant to the $\kappa$ field to have $<\kappa>=0$ over the 
area of $\tilde R\le 5$ (dashed), to have $<\kappa>=0$ at $\tilde R=5$ (dash-dotted). 
The solid lines are for the true mass distribution. It is seen that different
choice of the constant results significantly different enhancements. However, 
even in the case of $<\kappa>=0$ over the field within $\tilde R=5$, 
%which gives the lowest enhancement factor, $f\sim 10$ and $ 2$ at $\nu_t=4.5$ 
which gives the lowest enhancement factor, $f\sim 10$ and $ 6.2$ at $\nu_t=4.5$ 
for the isothermal and NFW clusters, respectively. 
%It has been argued that the 'dark clump' found near cluster A1942 is unlikely 
%to be a noise peak resulting from the chance alignment of source galaxies. This is based on the 
%peak statistics without considering the cluster environment, which gives
%the average number of noise peaks with $\nu\ge 4.5$ in the considered area of $R_c\le R\le 5R_c$
%to be about $0.01$. From our results shown here, the number of noise peaks is increased by 
%a factor of $f\sim 10$ around an isothermal cluster. This leads to an order of magnitude increase 
%in the probability that the seen 'dark clump' near A1942 is a false one without
%corresponding to real mass concentrations. 
%It is known that without considering the cluster environment,  
%the average number of noise peaks with $\nu\ge 4.5$ in the considered area of $R_c\le R\le 5R_c$
%is about $0.01$. Thus in the isothermal cluster case, the corresponding number of 
%peaks increases to $\sim 0.1$, which leads to an order of magnitude enhancement
%in the probability that the 'dark clump' found near cluster A1942 is a noise peak.

The strong association of noise peaks with their nearby
clusters can have significant effects on cosmological studies with  
weak lensing effects. It has been argued that the $^\prime$dark clump$^\prime$ 
found near cluster A1942 (e.g., Erben et al. 2000) is unlikely
to be a noise peak resulting from the chance alignment of source galaxies. This is based on the
peak statistics without considering the cluster environment, which gives
the average number of noise peaks with $\nu\ge 4.5$ in the considered area of $R_c\le R\le 5R_c$
to be about $0.01$ for $\theta_G=0.5\hbox{ arcmin}$. 
From our results shown here, the number of noise peaks can be increased by
a factor of $f\sim 10$ if the density profile of the cluster is approximately
isothermal. This leads to an order of magnitude increase
in the probability that the seen $^\prime$dark clump$^\prime$ near A1942 is a false one without
corresponding to real mass concentrations. Thus in order to properly quantify the statistical
significance of a $^\prime$dark clump$^\prime$ near a cluster, 
the cluster mass distribution has to be taken into account. 
%Our model provides a useful estimate for this. 
Specifically, in observational weak lensing studies, such as the one
for A1942, one can first make an approximate modeling of the mass distribution of the main cluster
from the weak lensing measurements. Then our model can be applied to study the 
occurrence probability of noise peaks in its surrounding area.
Such a probability is more appropriate than the one from the blind estimate without considering the 
existence of the cluster itself. It should also be emphasized that this 
probability depends sensitively on the density profile of the cluster.
Moreover, because of its strong dependence on the density profile 
of the cluster, the spatial distribution of noise peaks can mimic, to a certain extent, 
the true substructure distribution within the cluster.
Thus the noise peaks can contaminate the cosmological information inferred from 
weak lensing substructure analyses. 

%Future large weak lensing surveys are expected to be able to give rise to 
%large samples of weak-lensing-identified clusters of galaxies. Their abundances
%as well as their spatial distributions contain valuable cosmological information.
%For the abundance analyses, it has been well known that for low peaks, 
%the noise ones dominate over the true ones, making only the high peaks useful in cosmological studies.
%Our studies shown here further reveal that the spatial distribution of noise peaks
%is not random, and has close correlations with that of true clusters of galaxies.
%In fact, although not discussed here, it can be seen that in general noise peaks 
%correlate with large-scale mass distributions.
%These would bring errors in the correlation functions of weak lensing clusters.
%% Such correlations can cause excessive small-scale correlations for 
%%$\kappa$ peaks if the noise peaks cannot be marked out accurately. Furthermore, 
%We will present detailed studies on this aspect in one of our future papers.

%So far we have been analyzing the influence of clusters on noise peaks.
We now turn to the inner region with $R\le R_c$ or $\tilde R\le 1$. 
The highest peak within this region is assumed to correspond
to the signal from the cluster itself. We then analyze how the height and position of this
main-cluster peak are affected by the existence of noise. 

In the case without noise, the expected weak lensing peak heights should be
the same for a sample of clusters with the same mass distribution and at the same redshift. 
With noise included, however, their measured peak heights do not have the same value anymore. 
Instead, they follow a probability distribution. Based on our analyses shown in the previous section, 
the probability can be written as 
\begin{equation}
p(\nu)={n_{peak}(\nu)\over [\int n_{peak}(\nu^{'})d\nu^{'}]},
\end{equation}
%\begin{eqnarray}\label{w11}
%p_{pk}(\nu_0)=&&{1\over \int n_{pk}(\nu)d\nu} \exp \bigg [-{(K^1)^2+(K^2)^2\over \sigma_1^2}\bigg ]
%\bigg \{ {1\over 2\pi\theta_*^2}{1\over (2\pi)^{1/2}}\bigg\}
%\exp\bigg [-{1\over 2}\bigg ( \nu_0-{K\over \sigma_0}\bigg )^2\bigg ] \times \nonumber \\
%&&\int {dx\over [2\pi(1-\gamma^2)]^{1/2}}\exp\bigg \{-{ [{x+(K^{11}+K^{22})/ \sigma_2
%-\gamma(\nu_N-K/\sigma_0)}]\over 2(1-\gamma^2)}\bigg \} \times  F(x), \nonumber \\
%\end{eqnarray}
where $n_{peak}(\nu)$ is given by Eq.(15) and the central values of $K, K^1, K^2, K^{11}$
and $K^{22}$ at $\tilde R=0$ are used in calculating the probability.
In Figure 7, we show $p(\nu)$ 
%the probability distributions of the $\kappa$-peak height 
for the isothermal cluster (upper panel) and the NFW cluster (lower panel), respectively.
In each panel, the solid line is for the theoretical result from Eq. (31), and
the histograms are for the result from the $100$ Monte Carlo simulations.
For each simulation, we identify the highest peak within the region $\tilde R\le 1$
to be the one corresponding to the main cluster. 
The true cluster mass distribution is considered without additional treatments
for the mass-sheet degeneracy.
Firstly, it is seen that the solid line is in very good agreement with the results from simulations,
showing that Eq.(31) can properly model the effects of noise on the measured 
height of the main-cluster peak. Secondly, it is noted that the average value of the measured 
peak height is shifted toward larger $\nu$ with respect to the true height of the cluster
(indicated by the vertical dotted line in each panel). 
The specific value of the shift depends sensitively on the density profile of the cluster.
For the isothermal cluster, the shift is about $\Delta \nu\sim 1.1$ from $\nu=5.6$
of the true peak height to $\nu=6.7$, whereas for the NFW cluster, the shift is smaller 
%with $\Delta \nu\sim 0.5$. 
with $\Delta \nu\sim 0.8$. 
This systematic shift due to the existence of noise
is closely related to the fact that the differential peak number distribution of
a pure Gaussian noise field (considering only maxima peaks) is peaked at $\nu\sim 1$,
rather than at $\nu=0$.
Such a positive shift can increase the number of weak-lensing cluster
detections, and thus affect the corresponding cosmological applications considerably. 
In the next subsection, we present a detailed model to calculate quantitatively the
expected number of weak-lensing convergence peaks, true or false, from a large-scale 
weak lensing survey, taking into account such a shift and the enhancement of the noise 
peaks around clusters of galaxies. This model allows us to be able to use directly the high peaks 
detected in a large-scale weak-lensing convergence map as cosmological probes
without the need of expensive follow-up observations to differentiate true and false peaks.

The existence of noise also results a spatial offset for the identified main-cluster-peak
from the true central position of the cluster. This can be understood as follows. Without noise,
the main-cluster-peak in the weak lensing convergence map locates exactly at the
cluster center $(R=0)$ in our considered cases of spherical clusters. At this spatial location,
the conditions $\partial_i K=0$ $(i=1,2)$ are satisfied. With noise, however, 
the location of a peak should satisfy necessarily the conditions 
$\partial_i K_N=\partial_i (K+N)=0$ $(i=1,2)$, rather than $\partial_i K=0$ $(i=1,2)$. 
Given a realization of a noise field N and considering a small offset $\Delta R_i$ $(i=1,2)$, 
the original central peak now shifts to a location which can be estimated 
by $(\partial_{ij}K) \Delta R_j=-\partial_i N $ $(i,j=1,2)$,
where we have used $\partial_i K_N\approx \partial_{ij}K \Delta R_j$. Thus 
the offset distribution depends on the distribution of $\partial_i N$, which, in the considered
Gaussian case, is characterized by $\sigma_1$ defined in Eq. (7). It 
also depends sensitively on the density profile of the main cluster through $\partial_{ij}K$.
Specifically, the offset distribution can be approximately modeled by 
\begin{equation}
%p(\Delta R)d \Delta R={2\pi \Delta R d\Delta R \int n_{peak}(\Delta R,\nu)d\nu \over [\int_0^{R_c(R_s)}d\Delta R (2\pi \Delta R)
p(\Delta R)d \Delta R\approx {2\pi \Delta R d\Delta R \int n_{peak}(\Delta R,\nu)d\nu \over [\int_0^{R_c}d\Delta R (2\pi \Delta R)
\int n_{peak}( \Delta R,\nu)d\nu]},
\end{equation}
where the peak number density $n_{peak}(\Delta R,\nu)$ can be obtained from Eq. (15)
by Taylor expansion of the quantities of $K, K^i$ and $K^{ij}$ with respect to 
$\Delta R$. In our considered case, $\partial_i K(R=0)=0$ and
$\partial_i (K^{jk})(R=0)=0$, then to the linear order of $\Delta R$, 
we have 
\begin{equation}
p(\Delta R)d \Delta R\approx d \Delta R {\Delta R \exp\{-[K^{11}(0)\Delta R/\sigma_1]^2\}
%\over \int_0^{R_c(R_s)}
\over \int_0^{R_c}
d\Delta R\hbox{ } \Delta R \exp\{-[K^{11}(0)\Delta R/\sigma_1]^2\}]}.
\end{equation}
In terms of $\Delta \tilde R$, Eq.(33) can be written as 
\begin{equation}
p(\Delta \tilde R)d \Delta \tilde R\approx d \Delta \tilde R {\Delta \tilde R 
\exp[-(\Delta \tilde R/\tilde \sigma)^2]\over \{0.5{\tilde \sigma}^2[1-\exp(-{\tilde\sigma}^{-2})]\}}
\end{equation}
where $\tilde \sigma^{-1}=[K^{11}(0)/\sigma_2](\sigma_2/\sigma_1)(R_c)$.
%where $\tilde \sigma^{-1}=[K^{11}(0)/\sigma_2](\sigma_2/\sigma_1)(R_c)$ for the isothermal
%case and $\tilde \sigma^{-1}=[K^{11}(0)/\sigma_2](\sigma_2/\sigma_1)(R_s)$ for the NFW case. 
In our considerations with the Gaussian smoothing scale
$\theta_G=0.5\hbox{ arcmin}$ and $R_c=1.7 \hbox{ arcmin}$, we have 
%$[K^{11}(0)/\sigma_2]\approx 0.25$ and $1.2$, and the corresponding 
%$\tilde \sigma\approx 0.58$ and $0.122$ for the isothermal and NFW cases, respectively.
$[K^{11}(0)/\sigma_2]\approx 0.25$ and $0.68$, and the corresponding 
$\tilde \sigma\approx 0.58$ and $0.22$ for the isothermal and NFW cases, respectively.
We then expect a wider distribution of the offset for the isothermal case
than that of the NFW case. Figure 8 shows the results of $p(\Delta \tilde R)$.
The symbols are the simulation results with diamond and plus for the isothermal and NFW cases, 
respectively. The solid and the dash-dotted lines are the corresponding results calculated from 
Eq. (34). 
%It is seen that Eq. (34) can model the offset distribution reasonably well.
Because we use the linear expansion in $\Delta \tilde R$ in Eq. (34), we expect
some inaccuracies of Eq. (34) when $\Delta \tilde R$ is relatively large.
On the other hand, we see that Eq. (34) still gives a reasonable description 
about the offset distribution, especially in terms of the location of the maximum 
probability. 
%The distributions are distinctively different for the two clusters. 
For the isothermal case, the offset distribution is peaked around $\Delta \tilde R\approx 0.4$,
%and the $95\%$ cumulative probability occurs at $\Delta \tilde R\approx 0.9$.
and for the NFW case, it is peaked at $\Delta \tilde R\approx 0.2$. 
%and $95\%$ of the 
%offsets are smaller than $\Delta \tilde R\approx 0.4$ according to Eq.(34). 
The relatively flat
%For the NFW case, it is peaked at $\Delta \tilde R\approx 0.1$, and $95\%$ of the 
%offsets are smaller than $\Delta \tilde R\approx 0.2$. The relatively flat
density distribution of the cored isothermal cluster can therefore lead to 
an offset of order of $R_c$ for the main-cluster-peak identified from weak-lensing analyses.
This large offset may have significant effects on the cluster mass and the mass profile estimations
from weak-lensing observations. For the NFW cluster, the offset is smaller.
% with $\Delta R< 0.2R_s$ due to its sharp decline of the density profile. 

The above analyses are mainly done for the two clusters shown in Figure 2, in which, 
the NFW cluster has the same central $K$ and the same mass within $R\le 5R_c$ as 
those of the isothermal cluster. This leads to $R_s=5R_c$, or a concentration 
parameter $c=1$ for the NFW cluster. To further demonstrate the dependence 
of the peak statistics on the density profile of a cluster, we also carry out detailed analyses  
for two more NFW clusters with different concentrations and peak heights. The parameters are 
$R_s=1.8R_c$, and the central $K=1.8K(R_s=5R_c)$ for one cluster, 
and $R_s=R_c$, and the central $K=3K(R_s=5R_c)$ for the other. 
These parameters are chosen so that the clusters have about the same mass as the NFW cluster with 
$R_s=5R_c$ considered above. In Figure 9, we show the results for the three NFW clusters. 
Figure 9a presents the profiles of $K/\sigma_0$. Figure 9b and Figure 9c are
for the height distributions and spatial offset distributions of the main-cluster peak. 
In each panel, the solid, dashed, and dash-dotted lines are for the cases with $R_s=R_c$,
$R_s=1.8R_c$ and $R_s=5R_c$, respectively. The histograms in Figure 9b and symbols in 
Figure 9c are for the results from Monte Carlo simulations. The vertical lines in Figure 9b 
indicate the true cluster peak height (dotted) and the shifted peak height 
corresponding to the maximum of the distribution (dash-dotted) in each case.
The dependence of the results on the cluster profile is clearly seen. With the
same mass, the more centrally concentrated the cluster density distribution is, 
the less the main-cluster-peak properties are affected by noise. The 
peak height shift due to noise (Figure 9b) is $\Delta\nu \approx 0.2$, $0.4$, and 
$0.8$ for the $R_s=R_c$, $R_s=1.8R_c$ and $R_s=5R_c$, respectively. The corresponding
spatial offset distribution is peaked at $\Delta R\approx 0.03R_c$, $0.06R_c$ and $0.2R_c$ 
for the three cases. It is known from Eq. (31) and (34) that $\Delta \nu$ and $\Delta R$
are mainly determined by the central value of $|\partial_{11}K+\partial_{22}K|$ of a cluster.
A more centrally concentrated density profile leads to a larger value of 
$|\partial_{11}K+\partial_{22}K|$, and thus smaller $\Delta \nu$ and $\Delta R$.

It is possible to incorporate our model 
on the height and the position of the main-cluster peak [Eq. (31) and Eq. (34)] 
into observational weak-lensing analyses of cluster density profiles so that
the noise effects can be properly taken into account. The detailed implementations
will be explored in our future studies.

\subsection{Peak number statistics in large-scale weak-lensing convergence maps}

Weak-lensing cluster search and subsequently statistical studies are among the 
important cosmological applications of weak-lensing observations. It is seen from 
our discussions in \S 3.1 that noise due to 
intrinsic ellipticities of background galaxies results false peaks in the weak-lensing
convergence map and the noise peak distributions near clusters of galaxies
are boosted depending on the density profile of the clusters. 
It is also shown that the peak heights corresponding to clusters of galaxies are
affected by the existence of noise that generates large scatters as well as systematic shifts. 
These effects can contaminate the cosmological applications of weak-lensing cluster statistics 
considerably if they are not taken into account carefully. In this subsection, we present
a model to calculate the number distribution of peaks $n_{peak}(\nu)$
over a large area, where $n_{peak}(\nu)$ includes both true peaks corresponding to clusters
of galaxies and the false peaks from intrinsic ellipticities of background galaxies. 
It should be noted that the false peak distribution near clusters of galaxies depends
on the density profile of clusters, and thus also carries important cosmological information.   
The model then allows us to use directly the peaks detected in the large-scale 
reconstructed convergence map from weak-lensing observations as cosmological probes
without the need to differentiate true or false peaks with follow-up observations. 

In our model, the surface number density of convergence peaks is written as
\begin{equation}
n_{peak}(\nu)d\nu=n_{peak}^c(\nu)d\nu + n_{peak}^n(\nu)d\nu.
\end{equation}
Here $n_{peak}^c(\nu)$ calculates peaks within virial radii $R_{vir}$ of dark matter 
halos, which includes the peaks corresponding to halos themselves taking into account the 
modified peak heights due to noise, and the noise peaks with enhanced number distributions
near clusters. The term $n_{peak}^n(\nu)$ is for the noise peaks in the field area away from 
dark matter halos.
Specifically, we have
\begin{equation}
n_{peak}^c(\nu)=\int dz{dV(z)\over dzd\Omega}\int dM\hbox{ } n(M,z) f(\nu, M,z),
\end{equation}
where $dV(z)$ is the cosmological volume element at redshift $z$, $d\Omega$ is the solid angle
element, $n(M,z)$ is the mass function of dark matter halos given the mass $M$ and redshift $z$,
and 
\begin{equation}
f(\nu, M,z)=\int_0^{R_{vir}} dR \hbox{ }(2\pi R)\hbox{ }n_{peak}(\nu,M,z),
\end{equation}
with $n_{peak}(\nu,M,z)$ given by Eq. (15). The quantities related to the 
density profile of dark matter halos $K, K^{i}$, and $K^{ij}$ in Eq. (15) are computed assuming 
a NFW profile for a halo with its characteristic density $\rho_s$ and scale $r_s$ (or the 
concentration $c_{vir}$) determined by $M$ and $z$ (Navarro, Frenk \& White 1996). 

For $n_{peak}^n(\nu)$, it can be written as
\begin{equation}
n_{peak}^n(\nu)={1\over d\Omega}\bigg \{n_{ran}(\nu)\bigg [d\Omega-\int dz{dV(z)\over dz} \int dM\hbox{ } n(M,z) 
\hbox{ }(\pi R^2_{vir})\bigg]\bigg \},
\end{equation}
where $n_{ran}(\nu)$ is the differential surface number density of 
noise peaks without foreground clusters.

We should emphasize that Eq. (35) does not correspond to a simple and direct 
summation of the number of peaks corresponding to true dark matter halos
and the peaks from pure noise field. Rather,
the two terms in the right-hand side of Eq. (35)
represent peaks from different environments. The considered area ($1 \hbox{ deg}^2$ here) 
is divided into regions occupied by dark matter halos and the field region away
from halos. The first term counts for total number of peaks in the regions occupied 
by dark matter halos, which includes both the peaks of the halos
themselves and the noise peaks within the virial radii of the halos. The halos are
assumed to be randomly distributed spatially with their average abundance given by the 
mass function $n(M,z)$. The peak abundance within the virial radius of a dark matter
halo is given by $f$ calculated through Eq. (37) and Eq. (15), which takes into account the effect
of the halo mass distribution on noise peak statistics and the effect of noise on the
peak height of the halo itself. To a certain extent, the treatment of the first term
is similar to the model of halo occupation distribution for galaxies, where the peaks 
corresponding to halos can be analogous to central galaxies and the noise peaks can be 
analogous to satellite galaxies. The second term $n^n_{peak}$ calculates the number of peaks
in the field region away from dark matter halos, which is the product of 
of surface number density of pure noise peaks and the spatial area not occupied by 
dark matter halos [cf. Eq. (38)].

It should also be noted that to apply Eq. (35), we implicitly assume that
the true convergence peaks corresponding to real mass concentrations 
are linked to individual dark matter halos. This is a very reasonable assumption 
concerning peaks with $\nu>3$. On the other hand, for lower peaks, they can 
arise from the projection effects of large-scale structures without 
being associated with certain primary halos. To analyze these low peaks, it may be 
more appropriate to model the large-scale density field as a Gaussian random field $K_{LSS}$.
Then the noise effect on the statistics of these low peaks can be investigated by 
analyzing peaks in the total field $K_{LSS}+N$, which is also a Gaussian random field.
In fact, it is possible to separate the mass distribution in the universe into highly
nonlinear part $K$ which can be modeled by dark matter halos and the part associated 
with large-scale structures that can approximately be treated as a Gaussian random field $K_{LSS}$. 
By studying the peaks in the total field $K+K_{LSS}+N$, one can possibly obtain the peak
statistics over a wide range of peak height, from low to high. We will pursue
along this line in our future investigations. In this paper, we primarily concern 
high peaks with $\nu\ge 3$, and thus Eq. (35) is used in our analyses.

To validate our modeling of Eq. (35), we also analyze the 
convergence peak statistics from the ray tracing simulations by 
White \& Vale (2004). The simulations we use are for the flat 
$\Lambda$CDM model with $\Omega_M=0.296, h=0.7, \sigma_8=0.93$, and $n=1$, 
where $\Omega_M, h, \sigma_8$ and $n$ are the dimensionless matter density of the universe, 
the present Hubble constant in units of $100\hbox{ km/s/Mpc}$, the rms of the extrapolated 
linear density fluctuation over $8\hbox{ Mpc}h^{-1}$ and the power index of the initial
power spectrum of density fluctuations, respectively. 
In total, 16 simulated convergence maps, each with $1024\times 1024$ pixels corresponding to
a field of view of $3\times 3\hbox{ deg}^2$, are used in our analyses. The source redshift is
$z_s=1$. The details of the simulations can be found in White \& Vale (2004). For each 
simulated convergence map, we add in noise from intrinsic ellipticities 
following Hamana et al. (2004). Specifically, a random Gaussian noise is added to each pixel
with the variance given by 
\begin{equation}
\sigma^2_{pix}={\sigma_{\epsilon}^2\over 2}{1\over n_g \theta_{pix}^2},
\end{equation}
where $\sigma_{\epsilon}$ is the rms of the intrinsic ellipticity of source galaxies,
$n_g$ is the surface number density of source galaxies, and $\theta_{pix}$ is the 
pixel size of the simulated map. We take $\sigma_{\epsilon}=0.4$, $n_g=30\hbox{ arcmin}^{-2}$,
and $\theta_{pix}=180/1024 \hbox{ arcmin}$. We then apply a Gaussian smoothing to each 
noisy convergence map. Here we use the smoothing scale $\theta_G=1\hbox{ arcmin}$, which 
is presumably the most optimal smoothing scale in weak-lensing 
cluster detections (e.g., Hamana et al. 2004). 

In our model calculations, we use the same set of parameters as described above. 
The Sheth-Tormen mass function (1999) and the NFW density profile
for dark matter halos are applied.

In Figure 10, we show the surface number density $n_{peak}(\nu)/ \hbox {deg}^2$ 
from our theoretical calculations (solid line) in comparison with 
those from numerical simulations (upper histogram). 
The simulation results are the results averaged over $16$ of 
$3\times 3 \hbox{ deg}^2$ convergence maps, 
i.e., we count the peaks over the total area of $16\times 3\times 3=144\hbox{ deg}^2$, 
and then divide the total peak number by $144\hbox{ deg}^2$ to obtain the average
surface number density of the peaks. 
%We adopt the Sheth-Tormen mass function (1999) in our calculations.
We see that our model calculations agree with the simulation
results very well.
In the plot, we also show $n_{peak}(\nu)$ without 
considering the noise (dotted line), which is given by
\begin{equation}
n_{peak}(\nu)=\int dz{dV(z)\over dzd\Omega} \int dM\hbox{ } n(M,z) \hbox{ }\delta[M-M(z,\nu)],
\end{equation}
where $\delta[M-M(z,\nu)]$ is a $\delta$ function representing 
the fact that without noise, a dark matter halo with given $M$ and
$z$ has a determined peak height. The lower histograms are for the results from 
numerical simulations without adding noise in the convergence maps. It is seen that
this set of histograms is systematically higher than that of the dotted line,
indicating significant projection effects from large-scale structures.

Furthermore, to demonstrate the effects of noise
on the main-cluster-peak heights, we also show (long-dashed line) the results of $n_{peak}(\nu)$
for peaks associated with true dark matter halos but taking into account
the change of the main-cluster-peak heights due to noise (cf. Figure 7 and Figure 9b). 
This is given by
\begin{equation}
n_{peak}(\nu)=\int dz{dV(z)\over dzd\Omega} \int dM\hbox{ } n(M,z) \hbox{ }p_{peak}(\nu,M,z),
\end{equation}  
where $p_{peak}(\nu,M,z)$ is given by Eq. (31) with $K,K^{i}$, and $K^{ij}$ determined
by $M$ and $z$. Comparing the long-dashed line with the dotted line,
we see that at $\nu>3$, the number of peaks $n_{peak}(\nu)$
associated with real dark matter halos increases by a factor of $\sim 2$. 
This is clearly due to the positive shift of the peak height of a dark matter halo
from the effects of noise (see Figure 7 and Figure 9b), 
which boosts certain peaks of relatively small dark matter halos initially
with $\nu<3$ in the case without noise to $\nu>3$ with the presence of noise. 
This is highly relevant to cosmological studies
with the number of true weak-lensing peaks, i.e., confirmed peaks by follow-up observations. 
Without properly taking into account the noise effects on the peak heights, Eq. (40)
would normally be used to predict the abundance of weak-lensing peaks above a certain threshold.
By fitting such a prediction with the observed abundance of true weak-lensing peaks 
to constrain cosmological parameters, we would obtain a set of biased cosmological parameters. 
The detailed analyses of such bias effects on cosmological parameters 
will be presented in our forthcoming paper. It should also be pointed out 
that the differences between the long-dashed and solid lines
represent the number of noise peaks. These differences are much larger than those calculated from 
$n_{ran}(\nu)$ without considering the enhancement of noise peaks near dark matter halos, 
indicating the importance in taking into account such an enhancement in peak number calculations. 

The results shown in Figure 10 are directly comparable with the ones shown in Figure B1 of 
Hamana et al. (2004). By analyzing the results from numerical simulations, they present a 
fitting correction to the number of peaks taking into account the
noise effects (the solid line in their Figure B1). While in qualitative agreements,
our results match the simulation results better. More importantly,
our model [Eq. (35)-(38)] is based on statistical analyses of peaks 
near cluster of galaxies, but not a fitting model to the simulations. The excellent
agreement between our model prediction (the solid line in Figure 10) and that from simulations  
(the upper histogram) shows that our model can indeed give a reliable prediction about
the number of convergence peaks in weak-lensing analyses. 
%we are able to avoid expensive follow-up observations to find true peaks. 
Thus we can use directly the distribution 
of the total number of peaks, true and false ones, to probe cosmologies.
This advantage would greatly strengthen the cosmological applications of weak-lensing
peak statistics.

\section{Discussion}

In this paper, we analyze the peak statistics in weak-lensing convergence maps
taking into account the mutual influences between the mass distribution of clusters
of galaxies and noise from intrinsic ellipticities of background galaxies.
Concerning the main convergence peak corresponding to a cluster, 
the existence of noise affects both its height and its location,  
resulting scatters as well as a systematically
positive shift for its height and an offset distribution for its position. 
The effects are sensitive to the density profile of the cluster itself. In the 
two considered cases of the cored isothermal and the NFW clusters with the same mass and
the same central $K$, the flat density distribution of the isothermal cluster
leads to a relatively large shift of the peak height with $\Delta \nu \sim 1$ 
and a wide offset distribution of the peak location with $\Delta R$ extending to 
the core radius $R_c$. 
For the NFW cluster, its inner density profile changes more rapidly and
has a relatively large $|\partial_{11}K+\partial_{22}K|$ at the center. 
This leads to smaller $\Delta \nu$ and $\Delta R$, with $\Delta \nu\sim 0.8$ and 
$\Delta R$ peaked at $\Delta R\sim 0.2R_c$. We further analyze different NFW clusters
of the same mass but with different concentrations. It is shown that the weak-lensing 
convergence peak signal of a dark matter halo is affected less as its
concentration gets higher. We have $\Delta \nu\sim 0.2, 0.4,$ and $0.8$, 
and $\Delta R\sim 0.03R_c, 0.06R_c$ and $0.2R_c$, for
$R_s=R_c$, $R_s=1.8R_c$ and $R_s=5R_c$, respectively.

%For the NFW cluster, its density profile changes rapidly
%with large $\partial_{ij}K$ at the central region. Thus its main-cluster-peak is less affected
%by the noise than that of the isothermal cluster, with $\Delta \nu \sim 0.5$, 
%and $\Delta R \le 0.2 R_s$. 
%Such effects can have profound impact on cosmological 
%applications of weak-lensing cluster studies. 
For noise peaks, their occurrence probability   
is boosted near cluster of galaxies. 
In the isothermal case, there are on average about $2.4$ peaks with $\nu \ge 5$
(note that we have $\nu=5.6$ for the considered cluster without noise) in the central
region with $R\le R_c$. While the higher one is usually defined to be associated with 
the cluster itself, the other is a noise peak without any related real structures. 
This noise peak can be mistakenly identified as a large substructure and therefore 
can complicate the interpretation of the cluster mass distribution and further the 
formation of clusters. For the NFW cluster with $R_s=5R_c$,  
the central region is cleaner than that of the isothermal case,
and contains on average $\sim 1.6$ peaks with $\nu \ge 5$.
%There is typically clean with a well-defined single high peak corresponding to the main cluster. 
In outside regions around a cluster with $R>R_c$, 
high noise peaks can be confused with $^\prime$dark clumps$^\prime$, and thus lead to 
incorrect conclusions regarding the structure formation. We find that
depending on the density profile of a cluster, the average number of high noise peaks
in the regions $R_c< R \le 5R_c$ can be significantly enhanced. This is locally due to the
quantity $K(R)$ mainly. But it is important to realize that the 
$R$-dependence, i.e., the profile of $K(R)$, plays key roles resulting 
a net overall enhancement in the considered outside region, especially when
the mass-sheet degeneracy is taken into consideration. For the cored isothermal cluster,   
the enhancement factor $f$ ranges from $\sim 5$ to $\sim 55$ for the number of peaks with 
$\nu\ge 4$, depending on the treatment of the mass-sheet degeneracy. For $\nu\ge 5$, 
$f$ is in the range of $\sim 20-600$. For the NFW cluster, the enhancement is weaker 
than that of the isothermal cluster for high peaks, 
with $f\sim 3.6-55$ for $\nu\ge 4$, and $f\sim 11-440$ for $\nu\ge 5$. 
This enhancement is important in analyzing the statistical significance of 
$^\prime$dark clump$^\prime$ occurrence near clusters of galaxies.

Taking into account the effects of noise on the main-cluster-peak heights and
the enhancement of the number of noise peaks near dark matter halos, we propose 
a model incorporating the mass function of dark matter halos to calculate the 
statistical abundance of convergence peaks over large scales. The model prediction 
is in good agreement with that from numerical simulations. It should be pointed out 
that because of the mutual effects of the mass distribution of dark matter halos
and noise, the noise peak abundance also carries important cosmological information,
especially the information related to the density profile of dark matter halos.
Our model then provides a theoretical framework to use statistically the total number of convergence
peaks, including the false ones from noise, as cosmological probes. This facilitates 
the cosmological applications of weak-lensing peak statistics in saving the expensive 
follow-up observations to distinguish true and false peaks. Furthermore, the useful 
cosmological information contained in the noise peak distribution is also 
included.

Concerning high peaks with $\nu\ge 3-4$, our model can predict the total convergence peaks
very well because most of the high peaks are associated with massive virialized dark matter
halos (and the noise peaks around them). For relatively low peaks, the projection effects
of large-scale structures can be important, which can change the peak height of clusters 
as well as produce their own convergence
peaks without being associated with virialized clusters (e.g., Tang \& Fan 2005;
Marian et al. 2009, 2010; Maturi et al. 2009). To a certain extent, the total 
weak-lensing convergence can be written as $K_N=K+K_{LSS}+N$, where $K$ and $N$ 
are the respective contributions 
from virialized dark matter halos and the noise as discussed in this paper,
and $K_{LSS}$ denotes the large-scale structure contribution beyond dark matter halos.
On very large scales where density perturbations
are approximately linear, the effects of large-scale structures can be described by
a Gaussian random field reasonably well (Maturi et al. 2009). Under this Gaussian 
assumption, our current model on peak distribution near clusters of galaxies
can be extended to include the effects of large-scale
structures by analyzing the sum of the two Gaussian random fields $N+K_{LSS}$, which is 
also Gaussian, while treating the cluster $K$ as a background. Then Eq. (35) to Eq. (38) 
can be used to study large scale peak statistics, in which the Gaussian field $N$ is 
replaced with the total Gaussian field $N+K_{LSS}$. It should be pointed out that physical
correlations between $K$ and $K_{LSS}$ can exist, which may complicate the analyses
considerably. The above approach essentially divides the 
structures into two parts, virialized dark matter halos, and linear large-scale perturbations
described by a Gaussian random field. On scales in between where 
nonlinear effects are significant and the Gaussian approximation of $K_{LSS}$ is not appropriate, 
the modeling on $K_{LSS}$ deserves careful investigations. A possible way to study them
is to extend the mass function of dark matter halos to include such nonlinear
but non-virialized structures. Detailed investigations on the projection effects 
of large-scale structures, both linear and nonlinear, will be performed as one of our 
main future tasks. 

%The scatters and positive shifts generated by the noise on the measured peak heights of clusters
%of galaxies have direct consequences on cosmological studies from weak-lensing cluster
%abundances. Our theoretical analyses give rise to a kernel given in Eq. (33), which depends on
%the mass distribution of clusters. Together with the mass function of dark matter halos,
%one can then make a better prediction on weak-lensing cluster abundances for a given 
%cosmology, which in turn is important in constraining cosmological parameters from 
%future weak lensing observations. Recently the feasibility to constrain cosmologies
%using weak-lensing high peaks directly without linking to clusters of galaxies
%has been discussed (Wang, Haiman \& May 2009; Maturi et al. 2009; Kratochvil, Haiman \& May 2009).
%Our studies shown in this paper can be readily extended to such analyses. Specifically, 
%we need to investigate the association of noise peaks and large-scale mass distributions.
%As discussed above, it is expected that the general weak-lensing peak detection will be
%enhanced by including the noise in the analyses. In fact, such an enhancement has been seen in  
%Kratochvil et al. (2009).

Weak lensing effects hold great potentials in cosmological studies. With thorough investigations  
on different systematics, we expect that future weak-lensing observations would bring us
a greatly improved understanding about the dark side of the universe.

\acknowledgments 
We sincerely thank the referee for the comments and suggestions, which
help to improve the paper considerably. 
This research is supported in part by the NSFC of China under grants
10373001, 10533010 and 10773001, and the 973 program
No.2007CB815401. Zuhui Fan is very grateful for the hospitality of 
the Institute of Astronomy, University of
Cambridge where part of the paper is done. 
%%%%%%%%%%%%%%%%%%%%%%%%%%%%%%%%%%%%%%%%%%%%%%%%%%%%
%%%%%%%%%%%%%%%%%%%%%%%%%%%%%%%%%%%%%%%%%%

%\end{document}

\begin{figure}
\plotone{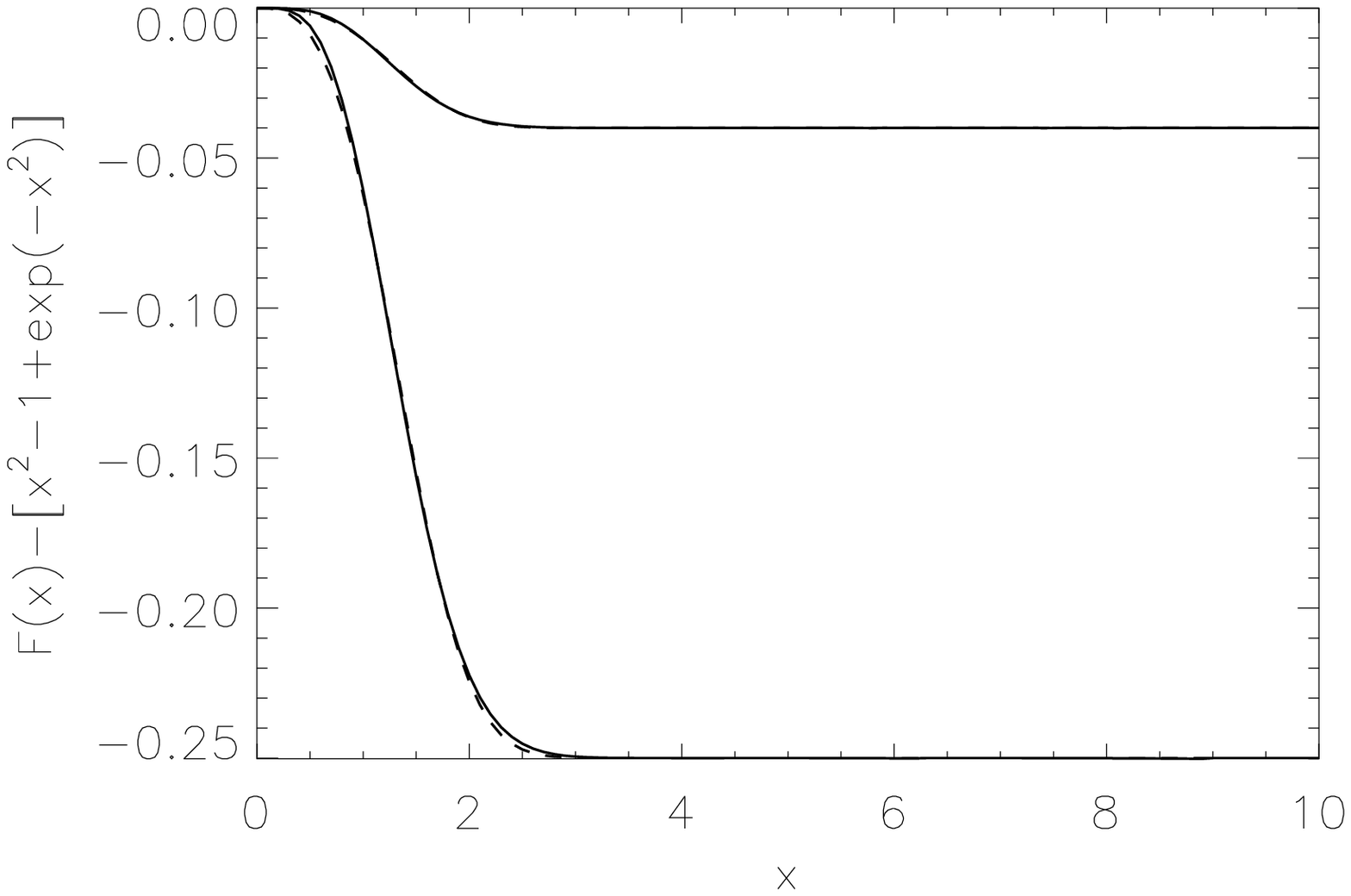}
%\plotone{fig1_ebinte_fit_new.ps}
\caption{The fitting results for $F(x)-[x^2-1+\exp(-x^2)]$ with $C_1=0.2$ (upper set)
and $C_1=0.5$ (lower set), respectively. For each set, the solid line is the
numerical result, and the dashed line is the fitting result from Eq.(17).
The $g$ values are $g(C_1=0.2)=3.3$ and $g(C_1=0.5)=3.5$, respectively. 
\label{yg6}}
\end{figure}

\begin{figure}
\plotone{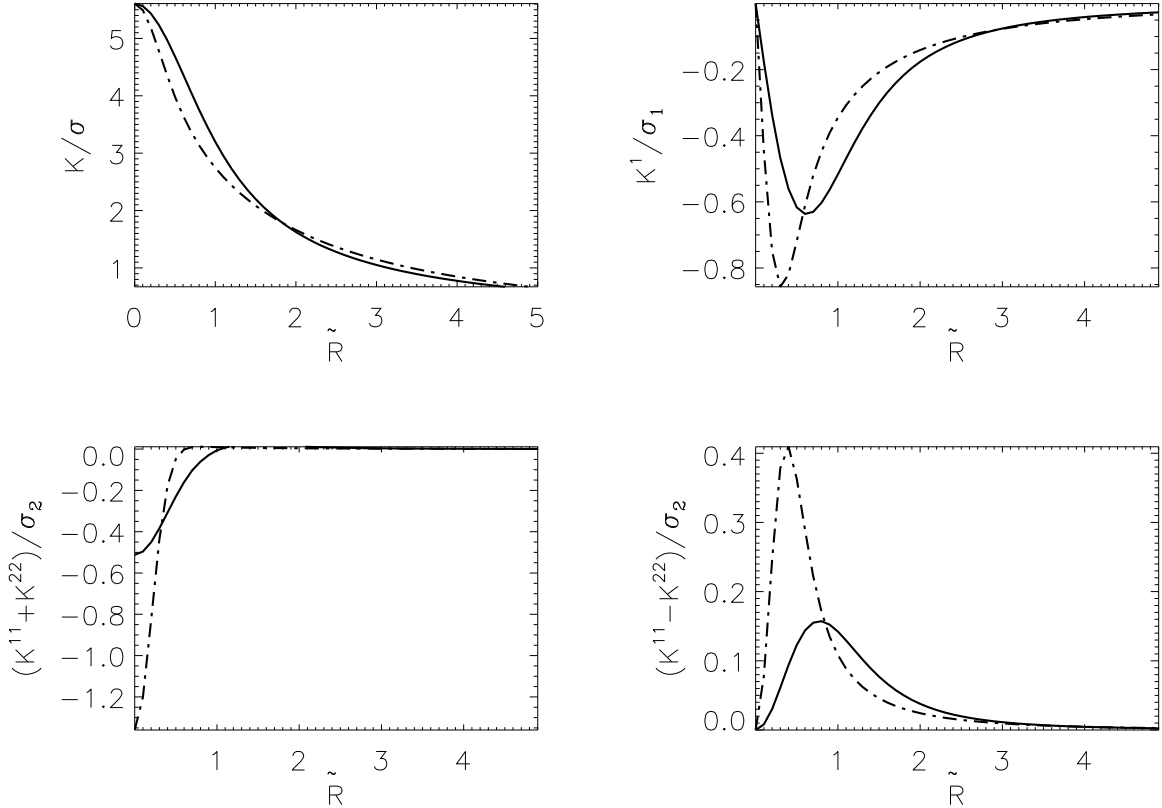}
%\plotone{fig2_smooth_rs5_iso.ps}
\caption{The profiles of $K/\sigma_0$, $K^1/\sigma_1$, $(K^{11}+K^{22})/\sigma_2$
and $(K^{11}-K^{22})/\sigma_2$ for the isothermal cluster (solid lines) and
the NFW cluster (dash-dotted lines), respectively. The two clusters have about the same 
mass and the same central $K/\sigma_0$. The Gaussian smoothing scale is
$\theta_G=0.5\hbox{ arcmin}$. The specific parameters for the isothermal cluster
is $R_c=1.7\hbox{ arcmin}$ and $\kappa_0/\sigma_0=6$ (the smoothed central 
$K/\sigma_0 \approx 5.6$). For the NFW cluster, $R_s=5R_c$, and the central 
$K/\sigma_0 \approx 5.6$. Here $\tilde R=R/R_c$. 
The profiles are calculated along $x_1$-axis with $x_2=0$.
\label{yg4}}
\end{figure}

\begin{figure}
\plottwo{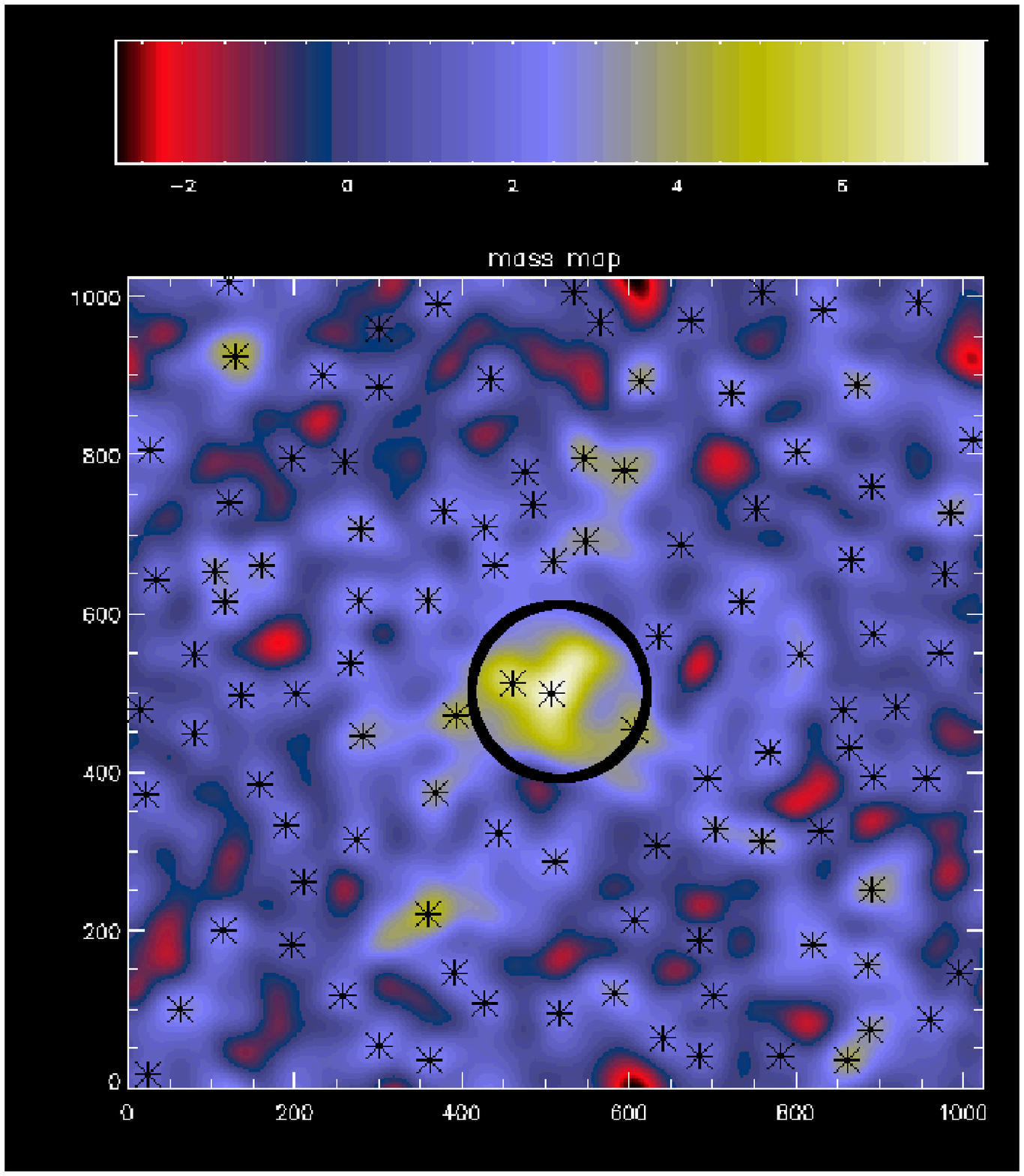}{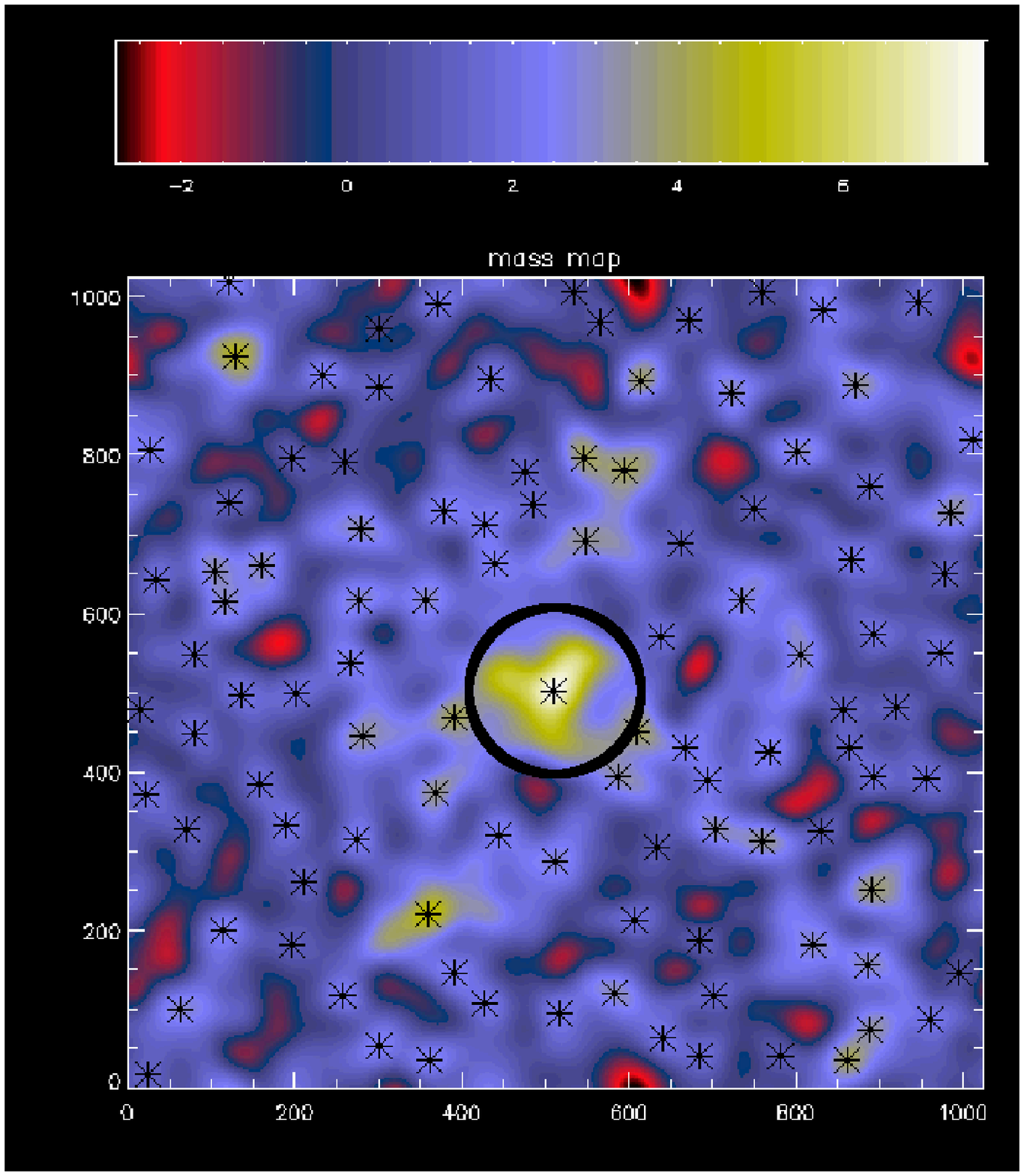}
%\plottwo{sis_s6_1.eps}{nfw_s6_rs5_1.eps}
\caption{The smoothed noisy convergence maps in terms of signal-to-noise ratio $K_N/\sigma_0$
for the isothermal cluster (Figure 3a, left) and
the NFW cluster (Figure 3b, right), respectively. The cluster parameters are the same as in 
Figure 2. The symbols show the positions of the peaks, and the circle indicates the 
%radius of $R_c$ ($R_s$). Each map has $1024^2$ pixels corresponding to the total area of 
radius of $R_c$. Each map has $1024^2$ pixels corresponding to the total area of 
$8.5\times 8.5 \hbox{ arcmin}^2$. 
\label{yg7}}
\end{figure}

\begin{figure}
\plotone{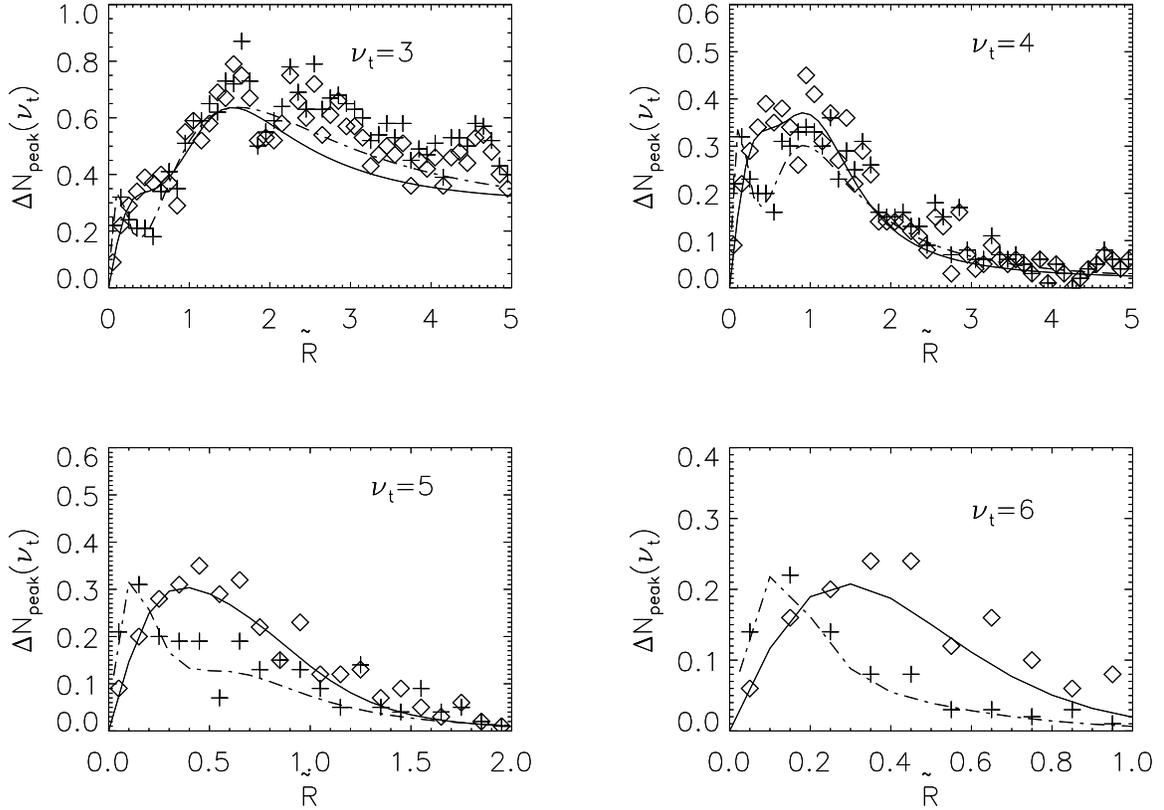}
%\plotone{fig_npeak_rdis.ps}
\caption{The radial distribution of the cumulative number of peaks 
%in the area within $5R_c(R_s)$
with $\nu\ge \nu_t$ around each cluster. The bin size is $0.1R_c$ ($\tilde R=0.1$). 
The solid, dash-dotted lines are the results for the
isothermal cluster and for the NFW cluster, respectively.
The diamond and the plus symbols are for the corresponding results averaged 
over $100$ Monte Carlo simulations. The cluster parameters are the same as those in Figure 2.
\label{yg7}}
\end{figure}

%\begin{figure}
%\plotone{f3.eps}
%%\plotone{fig3_allpeak.ps}
%\caption{The cumulative number distribution of noise peaks in the area within $5R_c(R_s)$
%around the cluster. The solid, dash-dotted, and dashed lines are the results for the 
%isothermal cluster, for the NFW cluster, and for the case without cluster, respectively.
%\label{yg7}}
%\end{figure}

\begin{figure}
%\plotone{f4.eps}
%\plottwo{fig4a_peak_cut12.ps}{fig4b_peak_cut12.ps}
%\plottwo{f4a.eps}{f4b.eps}
\plottwo{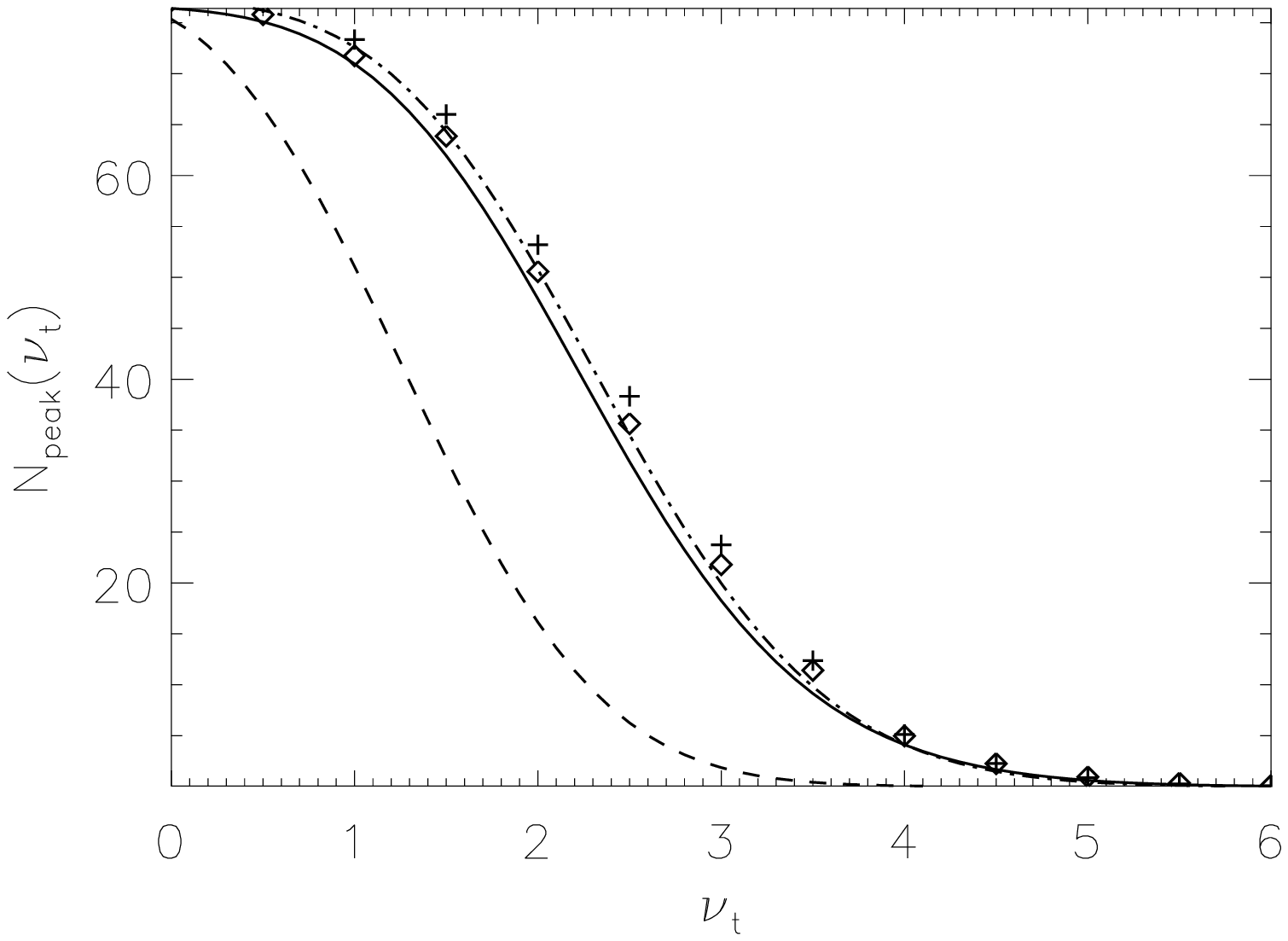}{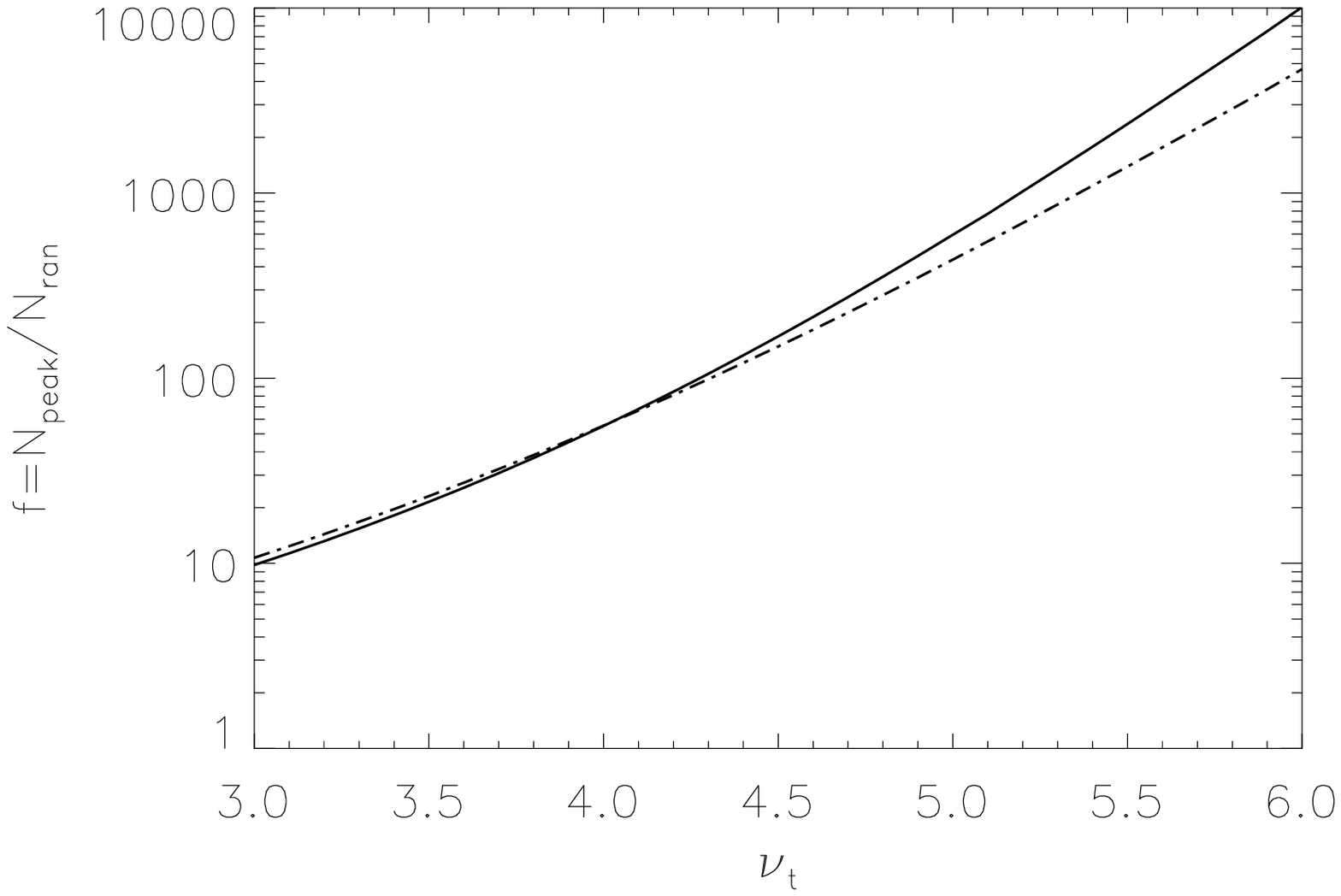}
%\plottwo{fig_cumu_outer.ps}{f4b.eps}
\caption{Fig.5a (left panel): The cumulative number distribution of noise peaks in the area within 
%$R_c(R_s)<R\le 5R_c(R_s)$. The solid, dash-dotted and dashed lines are for the isothermal, NFW, 
$R_c<R\le 5R_c$. The solid, dash-dotted and dashed lines are for the isothermal, NFW, 
and the pure noise cases, respectively.
The diamond and the plus symbols are for the average results 
from our Monte Carlo simulations of the isothermal cluster
and of the NFW cluster, respectively.
%Fig.5b (right panel): The ratio $f=N_{peak}/N_{ran}$ in the area within $R_c(R_s)<R\le 5R_c(R_s)$.
Fig.5b (right panel): The ratio $f=N_{peak}/N_{ran}$ in the area within $R_c<R\le 5R_c$.
The solid and the dash-dotted lines are for the isothermal and NFW clusters, respectively.
The cluster parameters are the same as those in Figure 2.
\label{yg8}}
\end{figure}

\begin{figure}
%\plotone{f5.eps}
%\plottwo{fig5a_peak_cut12.ps}{fig5b_peak_cut12.ps}
\plottwo{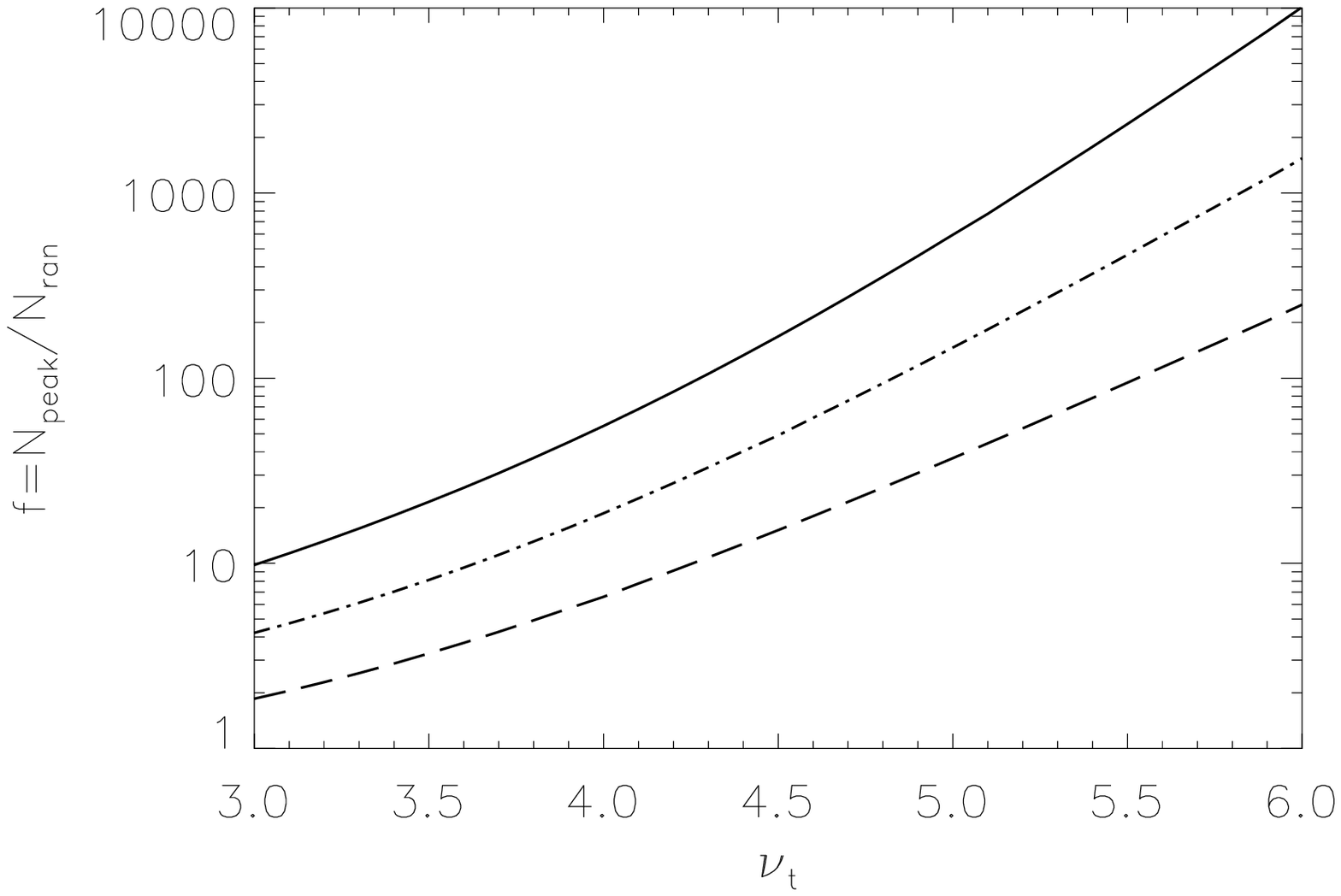}{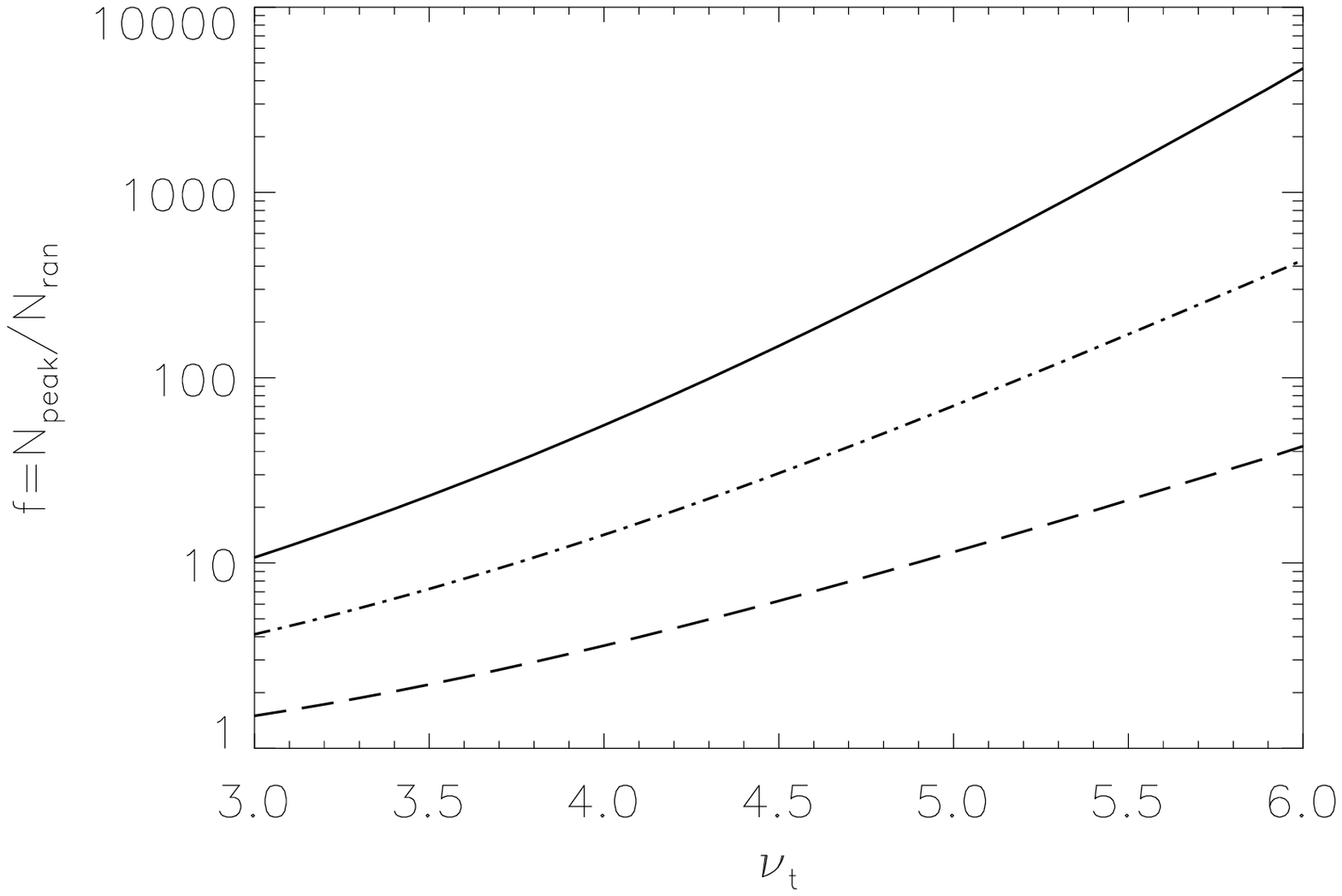}
%\caption{The ratio $f=N_{peak}/N_{ran}$ in the area within $R_c(R_s)<R\le 5R_c(R_s)$.
\caption{The ratio $f=N_{peak}/N_{ran}$ in the area within $R_c<R\le 5R_c$.
Fig.6a (left panel) is for the isothermal cluster, and Fig.6b (right panel) is
for the NFW cluster. In each panel, the solid, dash-dotted and dashed lines are for the
results with different treatments of the mass-sheet degeneracy, 
%specifically, with true cluster mass distribution, with $<\kappa>=0$ in the ring at $5R_c(R_s)$, 
%and with $<\kappa>=0$ within $5R_c(R_s)$, respectively. 
specifically, with true cluster mass distribution, with $<\kappa>=0$ in the ring at $5R_c$, 
and with $<\kappa>=0$ within $5R_c$, respectively. 
The cluster parameters are the same as those in Figure 2.
\label{yg10}}
\end{figure}

\begin{figure}
\plotone{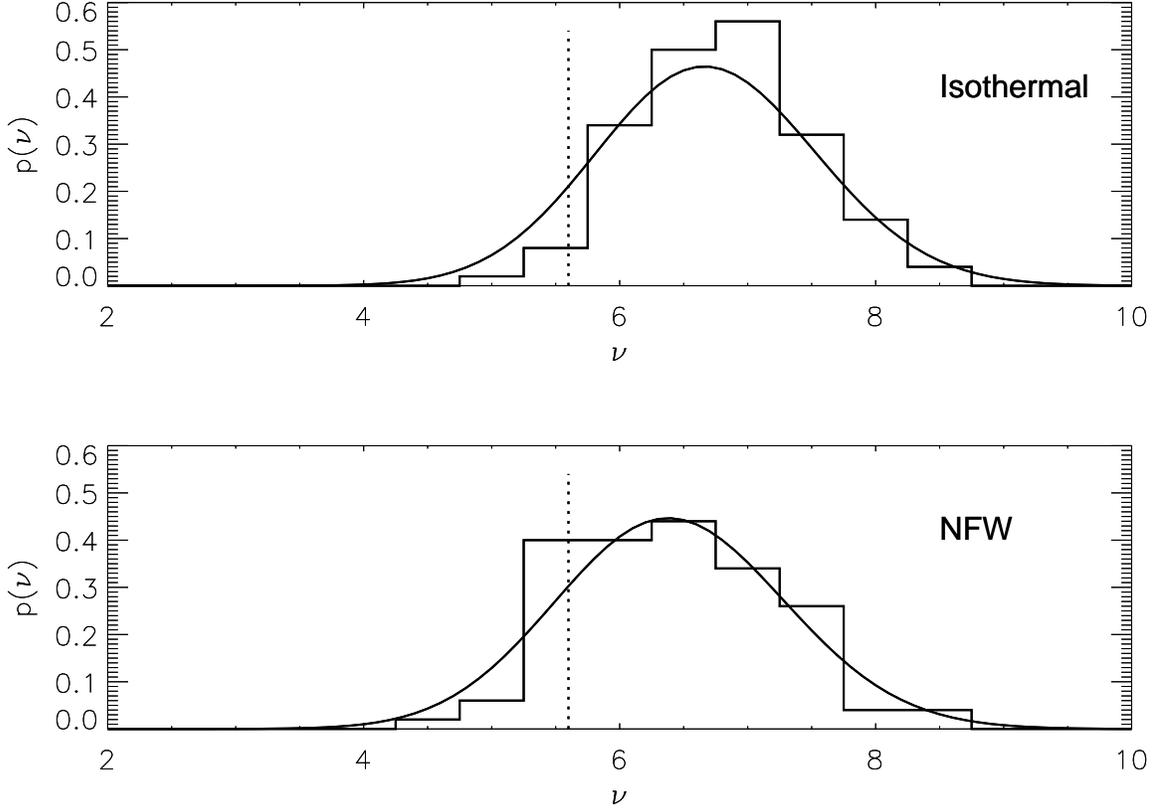} 
%\plotone{fig_peak_central_hdis_revise.ps}
%\plotone{fig6_peak_central.ps} 
\caption{The height distribution of the main-cluster-peak measured from 
noisy weak-lensing convergence maps.
The upper and the lower panels are for the isothermal and the NFW clusters, 
respectively. The solid line in each panel is for the result calculated from Eq. (31). 
The histogram is for the result from our Monte Carlo simulations. 
The vertical dotted line indicates the value of $\nu=5.6$, 
the true peak height of the considered cluster. 
The cluster parameters are the same as those in Figure 2.
\label{yg2}}
\end{figure}

\begin{figure}
\plotone{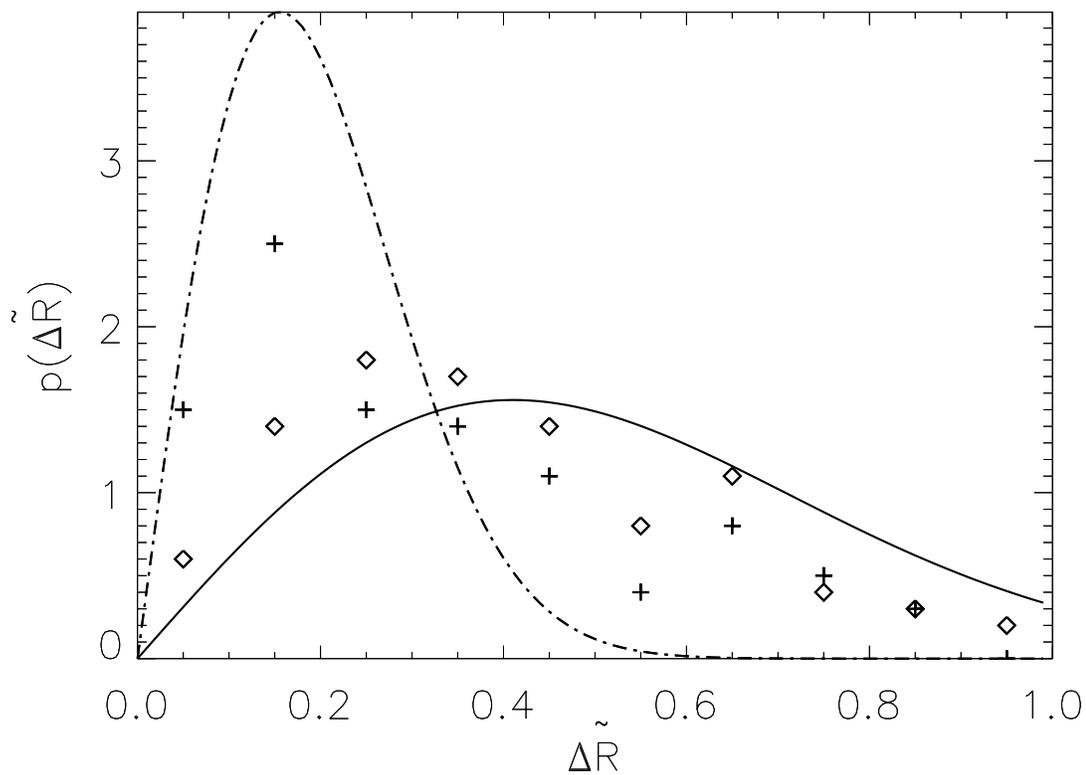}
%\plotone{fig_peak_central_rdis_revise_8jan10.ps}
%\plotone{fig6_peak_central.ps}
\caption{The spatial offset distribution of the measured main-cluster-peak position to the true 
center of the cluster. The solid and dash-dotted lines are for the results 
calculated from Eq. (34) for the isothermal cluster and the NFW cluster, respectively. 
The diamond and plus symbols are for the corresponding results from Monte Carlo simulations. 
The cluster parameters are the same as those in Figure 2.
\label{yg2}}
\end{figure}

\begin{figure}
%\plotone{f8.eps}
%\plotone{fig_peak_central_rdis_revise_8jan10.ps}
%\plotone{fig6_peak_central.ps}
\plottwo{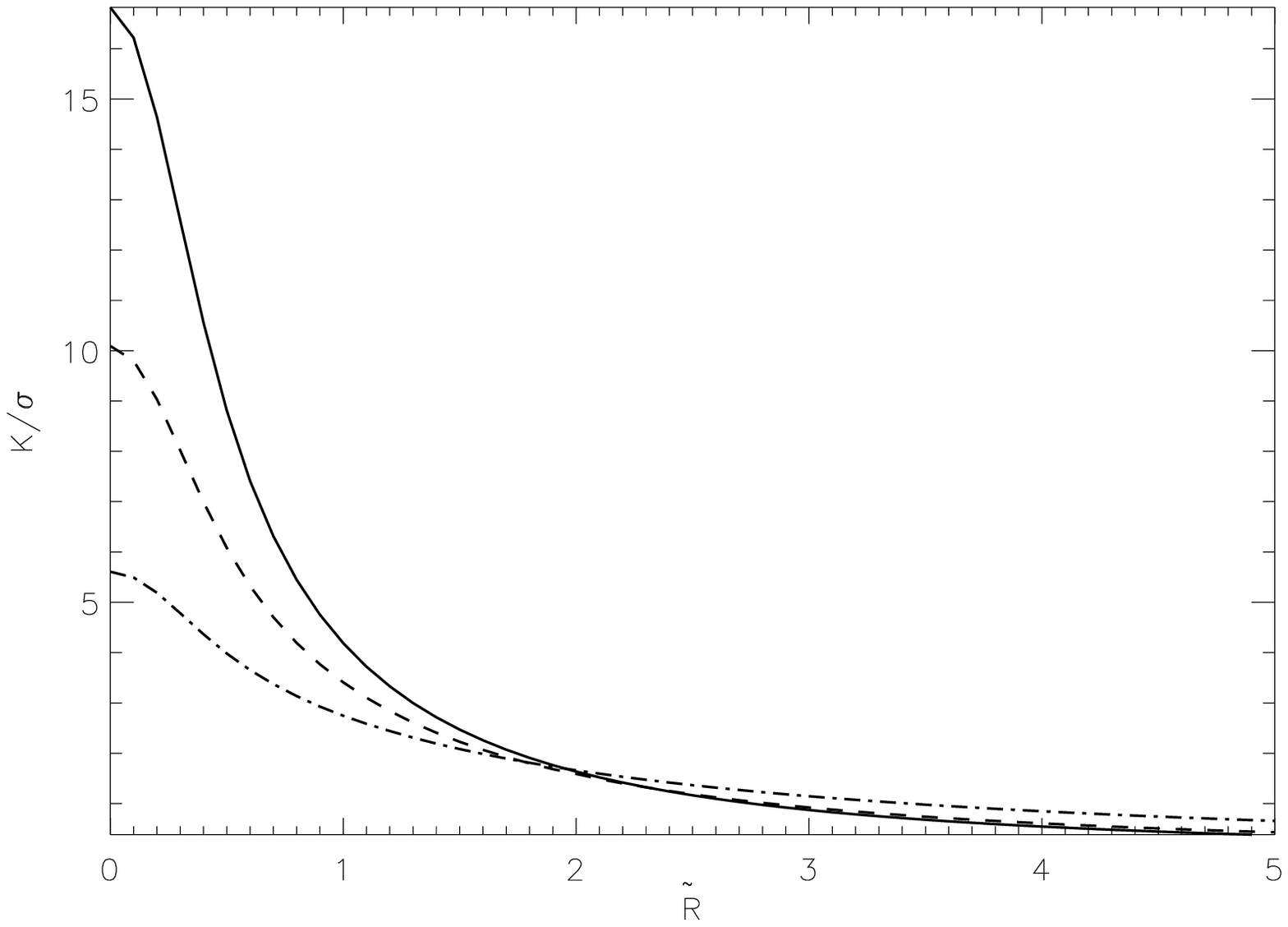}{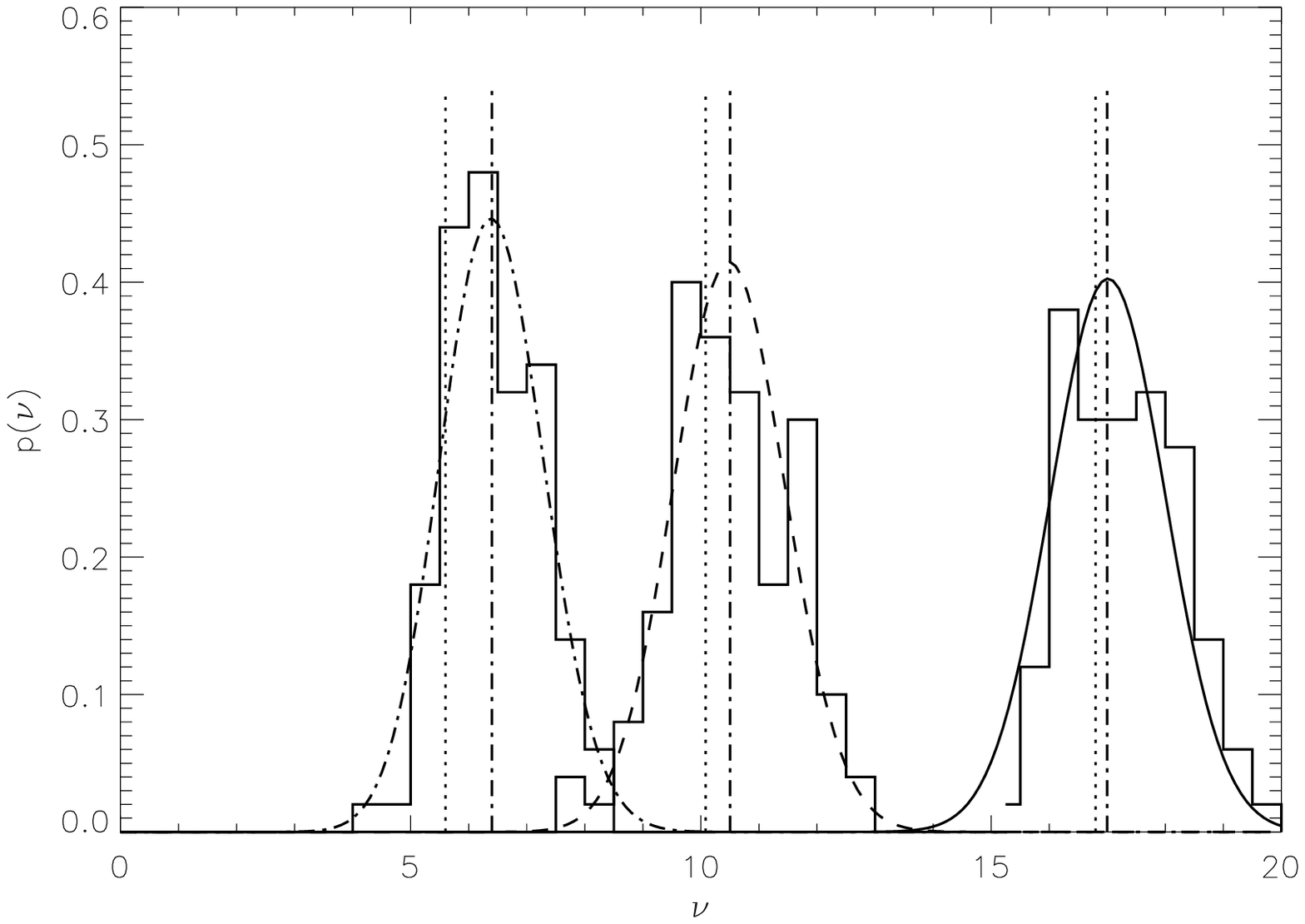}
\plotone{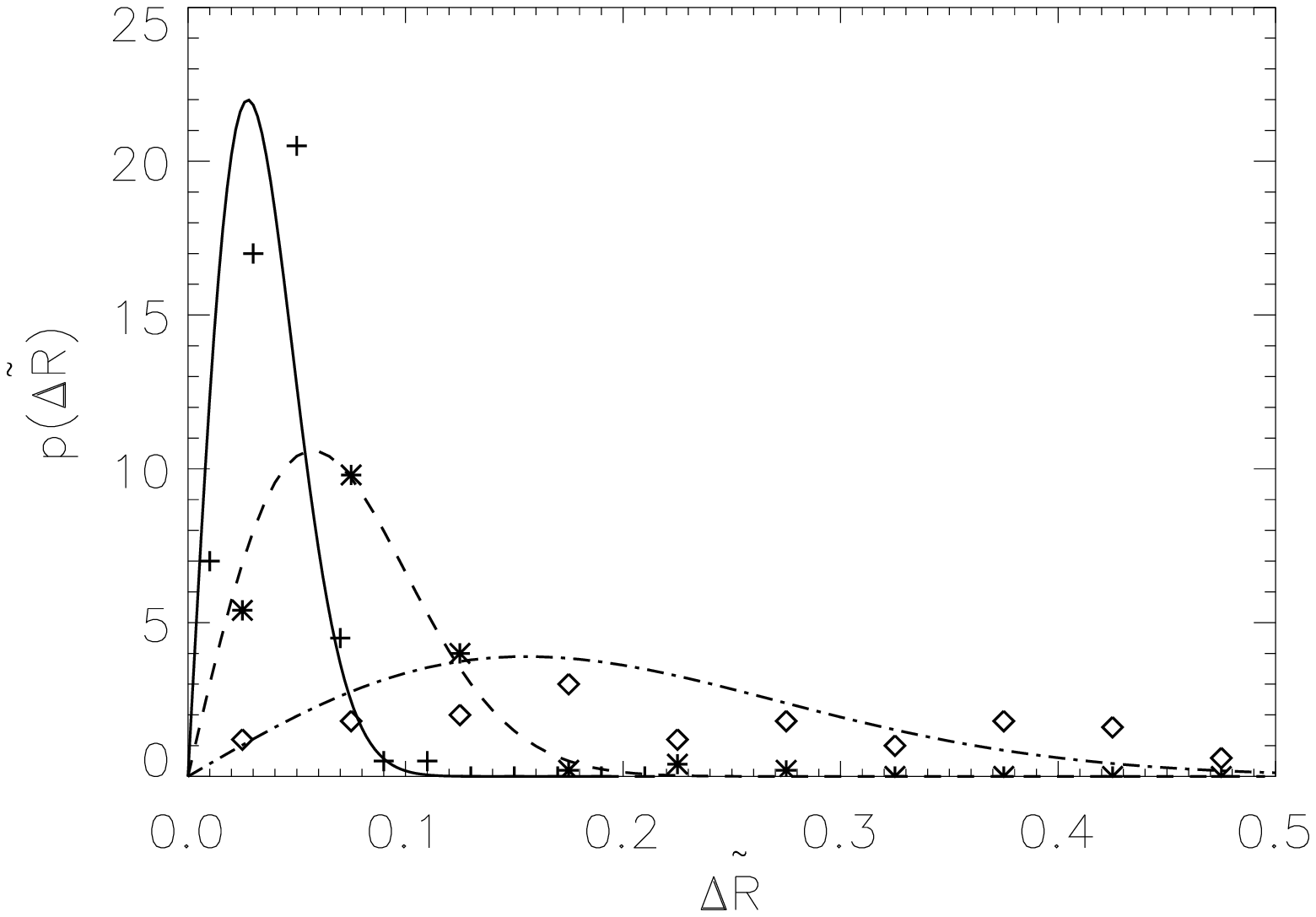}
\caption{Comparisons for three NFW clusters of about the
same mass but with different density profiles. The lines are the results from 
our theoretical calculations. The histograms and the symbols are the corresponding
results from Monte Carlo simulations. 
The solid line in each panel is for the cluster with $R_s=R_c$ and central $K/\sigma_0=3\times 5.6$.
The dashed line is for the cluster with $R_s=1.8R_c$ and central $K/\sigma_0=1.8\times 5.6$.
The dash-dotted line is for the cluster with $R_s=5R_c$ and central $K/\sigma_0=5.6$.
Fig. 9a (upper left panel) shows the profiles of $K/\sigma_0$ for the three clusters. 
Fig. 9b (upper right panel) is for the height distributions of the main-cluster-peak for the
three clusters, where the vertical dotted lines mark the values of the 
true peak height of the three clusters, and the vertical dash-dotted lines show
the $\nu$ value where $p(\nu)$ is peaked for each of the three clusters. 
Fig. 9c shows the spatial offset distributions of the main-cluster-peak position
for the three clusters.
\label{yg2}}
\end{figure}

\begin{figure}
%\plotone{f9.eps}
\plotone{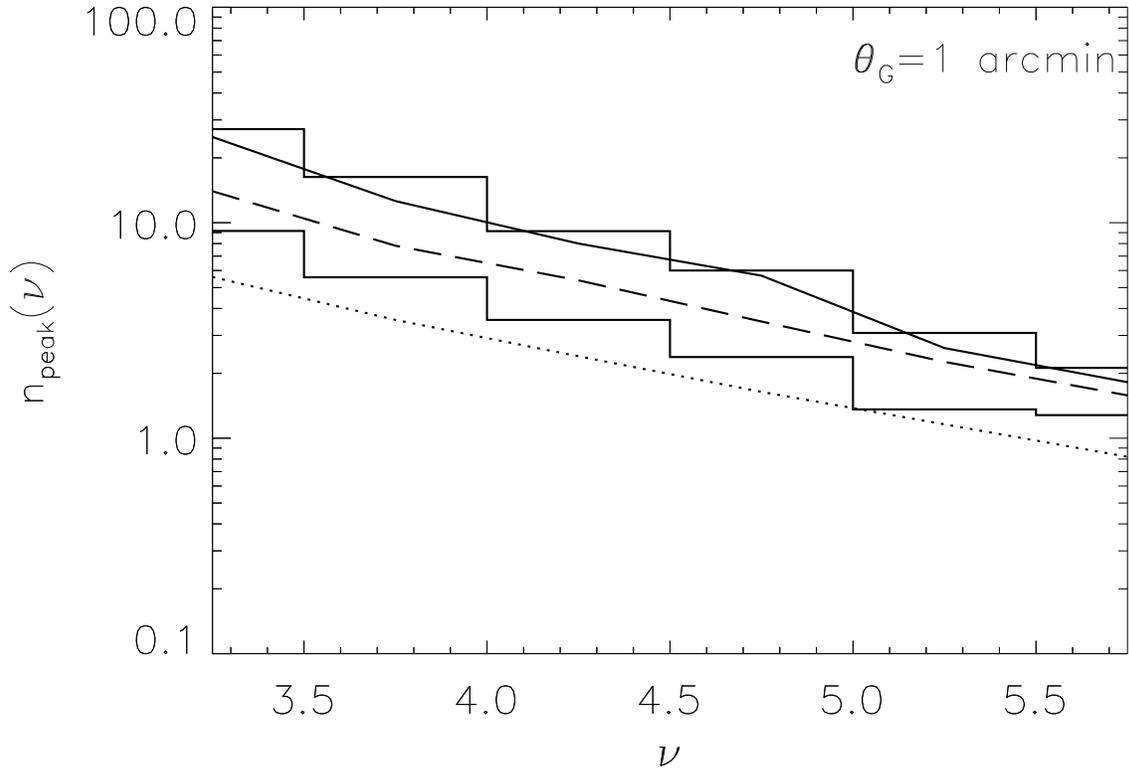}
%\plotone{fig_nfw_peak_theta_1_revise.ps}
%\plotone{fig6_peak_central.ps}
\caption{The differential peak height distribution in $1\hbox{ deg}^2$. The histograms are 
for the results averaged over $16$ ray-tracing simulations with $3\times 3 \hbox{ deg}^2$ each by 
White and Vale (2004). The lower and upper histograms correspond to the results 
without and with noise included. 
The dotted line is for the theoretical result considering only NFW halos without noise 
[Eq. (40)]. The long-dashed line is for the result of dark matter halos including 
the noise effects on the their peak heights calculated from Eq. (41) and Eq. (31). 
The solid line is for the prediction from Eq. (35)-Eq. (38) including the 
full noise effects. 
\label{yg2}}
\end{figure}

\end{document}